\documentclass[preprint,12pt]{elsarticle}
\usepackage{amssymb}
\usepackage{amsmath}
\usepackage{graphicx}
\usepackage{dcolumn}
\usepackage{bm}
\usepackage{amssymb}
\usepackage{color}
\usepackage{graphicx}
\usepackage{amsmath}
\usepackage{mathrsfs}
\usepackage{times}
\usepackage{subeqnarray}
\usepackage{cases}
\usepackage{bm}
\usepackage{diagbox}
\usepackage{booktabs}
\usepackage{bbm}
\usepackage[export]{adjustbox} 
\usepackage[table,xcdraw]{xcolor}
\usepackage{colortbl}
\usepackage{epstopdf}
\usepackage{algorithm}
\usepackage{algpseudocode}
\definecolor{tabcolor}{rgb}{.105,.410,.113}
\usepackage{array}
\newcolumntype{C}[1]{>{\centering\arraybackslash}p{#1}}
\usepackage{hyperref}
\hypersetup{colorlinks=true, citecolor=blue, urlcolor=blue, linkcolor=blue}
\journal{Chaos, Solitons \& Fractals}

\begin{document}

	\title{Emergence of cooperation: A reputation-modulated reinforcement learning}
	\author[a]{Chenyang Zhao}
	
	\author[b]{Jiqiang Zhang}
	
	\author[c]{Li Chen\corref{cor1}}

	\author[a]{Yong Zou\corref{cor2}}

	\cortext[cor1]{Corresponding author. \textit{E-mail address:} chenl@snnu.edu.cn (L. Chen).}
	\cortext[cor2]{Corresponding author. \textit{E-mail address:} yzou@phy.ecnu.edu.cn (Y. Zou).}
	\affiliation[a]{%
		organization={School of Physics, East China Normal University,},
		addressline={200241},
		city={Shanghai},
		country={P. R. China}
	}
		\affiliation[b]{%
		organization={School of Physics, Ningxia University,},
		addressline={750021},
		city={Yinchuan},
		country={P. R. China}
	}
		\affiliation[c]{%
		organization={School of Physics and Information Technology, Shaanxi Normal University,},
		addressline={710062},
		city={Xi'an},
		country={P. R. China}
	}

\begin{abstract}

Reputation is widely recognized as a key mechanism for sustaining cooperation. However, most existing game-theoretic models treat reputation primarily as an external factor that modulates payoffs, interaction structures, or strategy update rules. In many social contexts, though, reputation operates primarily as information -- it shapes how individuals interpret their own experiences and assess the behavior of others. To bridge this gap, we propose a spatial prisoner’s dilemma game grounded in the reinforcement learning paradigm, in which agents equipped with Q-learning integrate both individual and social information via a locally defined reputation metric to guide their decisions. Our results reveal that reputation-modulated learning significantly promotes the emergence of cooperative behavior, and we observe a discontinuous phase transition from full cooperation to full defection as the temptation increases. Cooperation spreads through the nucleation of cooperative clusters, whereas the disintegration of these clusters drives the system into an absorbing state of complete defection. Overall, this study demonstrates that reputation facilitates cooperation not only by providing direct incentives but also by reshaping the social information landscape that agents rely on for learning and adaptation.
\end{abstract}
\date{\today }
\maketitle
	
\section{Introduction}\label{sec1}
Understanding the emergence of cooperation among self-interested individuals remains a central challenge in evolutionary game theory and complex systems research. As a prototypical model, the prisoner’s dilemma (PD) captures the fundamental tension between individual rationality and collective welfare: although mutual cooperation yields socially optimal outcomes, defection remains the dominant strategy for rational individuals \cite{Nielsen1985cooperation,Colman1995Theory,Nowak2006Evolution}.

To account for the pervasive cooperation observed in human and animal societies, a rich array of mechanisms has been proposed over the past decades, including kin selection \cite{Dawkins2006selfish, Wilson1975Sociobiology}, group selection \cite{Smith1964Group,Charlesworth2000Levels}, direct reciprocity \cite{Nowak2008Repeated}, indirect reciprocity \cite{Perc2013Interdependent,Nowak1998indirect, Ohtsuki2006indirect}, network reciprocity \cite{Nowak1992spatial,Szabo1998Evolutionary,Wang2013Interdependent,Szolnoki2010Dynamically}, dynamical reciprocity \cite{liang2022dynamical}, and reputation-based mechanisms \cite{xia2023reputation}. In addition, cooperation can be reinforced by institutional incentives such as reward and punishment \cite{Fehr2002punishment,Herrmann2008Antisocial,Clutton1995Punishment,Diekmann2015Punitive}, or shaped by environmental feedback \cite{tilman2020evolutionary,Joshua2016An}.

Among these, reputation has emerged as a particularly powerful driver of cooperation, largely through the lens of indirect reciprocity \cite{milinski2002reputation,nowak1998evolution,Ohtsuki2006The,panchanathan2004indirect,murase2023indirect}. Reputation is commonly classified according to the order of its evaluation rules. Under first-order norms, an individual’s reputation is determined solely by their own actions \cite{Liu2017Sustainable}; second-order norms additionally take into account the reputation of the interaction partner \cite{DONG2019Cooperation,HAN2022Role}; and third-order norms further incorporate the reputational states of both parties as well as the broader social context in which the interaction occurs, thus enabling increasingly sophisticated forms of indirect reciprocity \cite{YANG2019Evolution,schmid2023quantitative}.

Recently, reinforcement learning (RL) has emerged as a powerful paradigm for modeling human decision-making, offering a compelling alternative to imitation learning (IL), which has been widely adopted in previous game-theoretic studies~\cite{zheng2026brief}. Unlike IL, where individuals adjust their strategies by observing and copying others' behaviors, RL agents continuously update action values based on environmental feedback and refine their strategies through trial-and-error learning. Growing evidence suggests that the RL framework can account for a range of collective phenomena related to cooperation~\cite{TomovMulti2021,ZhangOscillatory2020,Wang2022Levy,Ding2023Emergence,zheng2024evolution,JiaLocal2021, ZhaoEvolution2025}, as well as the emergence of other important social norms~\cite{Zheng2024decoding,zheng2025decoding,zheng2025optimal}.

Within both IL and RL paradigms, an increasing body of work has sought to integrate reputation mechanisms into learning-based evolutionary dynamics. Existing studies indicate that reputation can effectively facilitate cooperation in IL settings~\cite{Quan2020Reputation,QUAN2021Reputation,QUAN2024Reputation,Feng2024An, zhang2024Reputation} as well as in RL frameworks~\cite{Zou2024Incorporating,ren2023reputation}. For instance, Hu et al.~\cite{Hu2024Reputation} introduced a reputation-based incentive mechanism under public supervision, where high-reputation individuals receive additional rewards, thereby reinforcing cooperative behavior. Fang et al.~\cite{Fang2025Evolution} proposed a payoff-driven reputation mechanism in which individuals base migration decisions on reputation information, and found that cooperation can persist even under low population densities.

Despite these advancements, most existing approaches treat reputation as an external instrument that directly modifies payoffs, interaction patterns, mobility choices, or strategy update rules. Comparatively little attention has been paid to how reputation influences the learning process itself. In reality, reputation can function not only as a behavioral incentive but also as a social signal that shapes how individuals interpret and integrate information from both their personal experiences and their social surroundings~\cite{cuesta2015reputation,molleman2014consistent,YANG2024Interaction}.

Motivated by this gap, we propose a reputation-modulated RL framework in which reputation dynamically regulates the relative weighting of different information sources during learning. Importantly, our model does not alter payoff structures or prescribe explicit behavioral rules; instead, reputation governs how agents integrate personal experience with socially acquired information when updating action values. Specifically, agents with low reputation rely predominantly on their own experience, whereas those with high reputation assign greater weight to information obtained from neighbors. This design captures a key feature of real social systems: individuals do not simply imitate others or mechanically pursue collective welfare. Rather, they interpret collective performance as a social signal that reflects environmental quality and long-term strategic viability, which in turn shapes their learning and decision-making processes~\cite{Lamba2014Social,molleman2014consistent}.

\begin{figure}[t]
\centering
\includegraphics[width=0.99\linewidth]{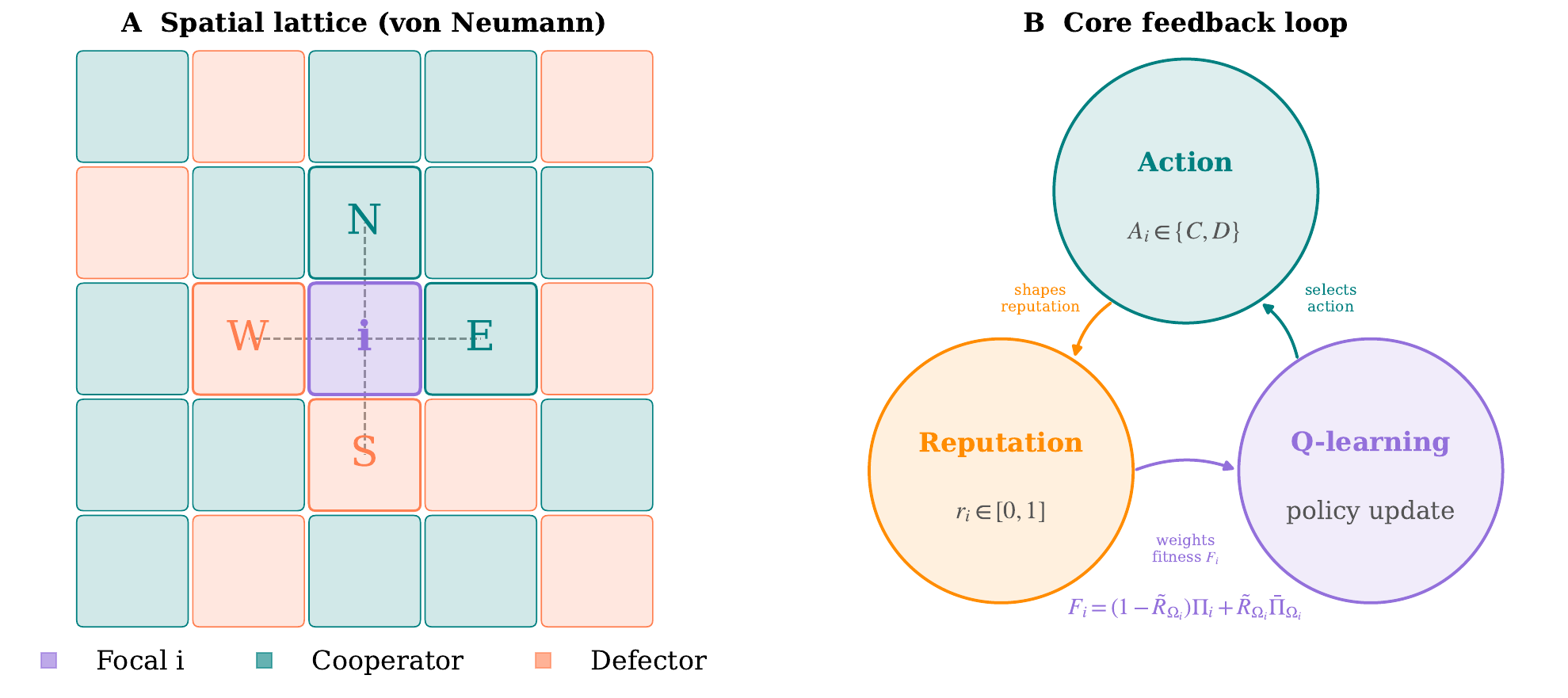}
\caption{\textbf{Schematic illustration of the reputation-modulated Q-learning model.}
Agents are placed on a square lattice and play the prisoner's dilemma game with their four nearest neighbors. After each round, an agent's reputation increases if it cooperates; otherwise, it decreases if it defects. The local reputation level determines how agents evaluate their learning reward: when neighborhood reputation is low, agents mainly rely on their own payoff, while a high reputation increases the weight of the neighbors' average payoff. The resulting reputation-weighted fitness is then used to update the Q table, which in turn guides subsequent action choices through an $\varepsilon$-greedy rule.}
		
		\label{fig:scheme}
	\end{figure}

In this work, we investigate how reputation-mediated social learning influences the evolution of cooperation (see Fig.~\ref{fig:scheme}). Our results show that reputation feedback significantly expands the cooperative region and induces a discontinuous transition between cooperative and defective states. Near the critical regime, the system exhibits pronounced bistability, marked by the coexistence of cooperative and defective attractors. Through analysis of the emergent Q-table structure and the spatiotemporal evolution of cooperative domains, we further uncover a nucleation mechanism underlying the formation of large-scale cooperation. Collectively, these findings demonstrate that reputation promotes cooperation not by directly incentivizing cooperative acts, but by reshaping the way social information is incorporated into the learning process.

\section{Model}\label{sec2}

We consider a population of $N$ agents distributed on a $L \times L$ square lattice network with periodic boundary conditions, where each agent interacts exclusively with its four nearest neighbors (the von Neumann neighborhood). At each discrete time step $t$, agents engage in repeated prisoner’s dilemma (PD) games and dynamically adapt their strategies via multi-agent reinforcement learning. 
	
At step $t$, agent $i$ selects an action $a_i(t) \in \mathbb{A} = \{C, D\} \equiv \{1, 0\}$. The interaction between agent $i$ and its neighbor $j$ yields a payoff determined by the payoff matrix:
	\begin{equation}
		\begin{pmatrix}
			\Pi_{CC} & \Pi_{CD}\\
			\Pi_{DC} & \Pi_{DD}
		\end{pmatrix} = \begin{pmatrix}
			R & S\\
			T & P
		\end{pmatrix} = \begin{pmatrix}
			1 & -b\\
			1+b & 0
		\end{pmatrix},
		\label{eq:matrix}
	\end{equation}

where the temptation to defect $T$ is $b$ ($0 < b \le 1$). This parametrization strictly preserves the basic inequalities $T > R > P > S$ and $2R > T + S$. The immediate accumulated payoff of agent $i$ from interacting with all its neighbors is defined as:
	\begin{equation}
	\begin{aligned}
		\overline{\Pi_{i}}(t) = \frac{1}{4} \sum_{j \in \Omega_i} \Big[ R a_i a_j + S a_i(1-a_j) \\
		+ T (1-a_i)a_j + P (1-a_i)(1-a_j) \Big],
	\end{aligned}
		\label{eq:reward}
	\end{equation}
	where $\Omega_i$ denotes the set of the four nearest neighbors of the focal agent $i$.

\begin{algorithm}[htp]
	\caption{Reputation-Modulated Q-Learning in Spatial Prisoner's Dilemma}
	\label{alg:rq}
	\begin{algorithmic}[1]
		\State \textbf{Input:} Lattice size $L$ ($N = L^2$), total Monte Carlo steps $T_{\rm MCS}$, parameters $b, c, c_{\max}, \varepsilon, \alpha, \gamma$.
		\State \textbf{Output:} Stationary cooperation level $\rho_C$.
		
		\State \textbf{Initialization:} For all agents $i \in \{1, \dots, N\}$, randomly assign initial strategy $a_i \in \{C, D\}$, initialize reputation $R_i \in [0, R_{\max}]$, initialize Q-table, and determine initial state $s_i \in \{0, \dots, 5\}$.
		
		\For{$t = 1$ to $T_{\rm MCS}$}
		\For{each agent $i$} 
		\State {Action selection via $\varepsilon$-greedy}
		\If{$\text{rand}() < \varepsilon$}
		\State $a_i \leftarrow$ random action in $\{0, 1\}$
		\Else
		\State $a_i \leftarrow \arg\max_{a} Q_i(s_i, a)$
		\EndIf
		\EndFor
		
		\For{each agent $i$} 
		\State Compute payoff $\pi_i$ by Eq.~\eqref{eq:reward}
		\State Compute fitness $F_i$ by Eq.~\eqref{eq:fitness}
		\State Update reputation $R_i$ by Eq.~\eqref{eq:update_repu} and clip it to[0, $R_{\max}$]
		\EndFor
		
		\For{each agent $i$} 
		\State Determine next state $s_i'$
		\State Update Q-table by Eq.~\eqref{eq:bellman}
		\EndFor
		
		\State $s_i \leftarrow s_i'$ for all $i$ 
		\EndFor
	\end{algorithmic}
\end{algorithm}

Unlike previous evolutionary models driven by objective material payoffs \cite{Feng2024An, Zou2024Incorporating}, here we introduce an endogenous social evaluation mechanism, where an agent's objective function (fitness) is modulated by the local social environment. Let $R_i(t) \in [0, R_{max}]$ denote the reputation value of agent $i$. The local environment's trustworthiness is quantified by the average reputation of agent $i$'s neighborhood:
	\begin{equation}
		\overline{R}_{\Omega_i}(t) = \frac{1}{4} \sum_{j \in \Omega_i} R_j(t).
	\end{equation}
The effective fitness $F_i(t)$ of agent $i$ that serves as the reward for RL is defined as a convex combination of personal and neighborhood payoffs:

	\begin{equation}
		F_i(t) = \left( 1 - \overline{R}_{\Omega_i}(t) \right) \overline{\Pi_i}(t) + \overline{R}_{\Omega_i}(t) \overline{\Pi}_{\Omega_i}(t),
		\label{eq:fitness}
	\end{equation}
where $\overline{\Pi}_{\Omega_i}(t) = \frac{1}{4} \sum_{j \in \Omega_i} \overline{\Pi_j}(t)$ is the average payoff of the neighborhood, and $\overline{R}_{\Omega_i}(t) = \overline{R}_{\Omega_i}(t) / R_{max}$ is the normalized reputation weight. 
 
Without loss of generality, we fix $R_{\max}=1$ throughout this work. Crucially, when the local reputation is negligible $\overline{R}_{\Omega_i}\rightarrow 0$, the agent behaves as a \emph{homo economicus}, focusing solely on its own payoff $\overline{\Pi_{i}}$. Conversely, a high-reputation environment $\overline{R}_{\Omega_i}\rightarrow 1$ triggers an ``enlightened self-interest'' heuristic, compelling the agent to internalize the collective payoff.

    \begin{table}[htp]
    \begin{tabular}{c|cc}
    \arrayrulecolor{tabcolor}\toprule [1.4pt]
    \hline
    \diagbox{State}{Action}& C ($a_1$) &  D ($a_2$) \\
    \midrule [0.5pt]
    \hline
    0 $(s_{0})$ & $Q_{s_{0},a_{1}}$ & $Q_{s_{0},a_{2}}$  \\
    1 $(s_{1})$ & $Q_{s_{1},a_{1}}$ & $Q_{s_{1},a_{2}}$ \\
    2 $(s_{2})$ & $Q_{s_{2},a_{1}}$ & $Q_{s_{2},a_{2}}$ \\
    3 $(s_{3})$ & $Q_{s_{3},a_{1}}$ & $Q_{s_{3},a_{2}}$\\   
    4 $(s_{4})$ & $Q_{s_{4},a_{1}}$ & $Q_{s_{4},a_{2}}$  \\  
    5 $(s_{5})$ & $Q_{s_{5},a_{1}}$ & $Q_{s_{5},a_{2}}$  \\  
    \hline
    \bottomrule[1.4pt]
    \end{tabular}
    \caption{Q-table for each Q-learning player. The state $s_{0,...,5}$ corresponds to the number of cooperators in its neighborhood, i.e. the four nearest neighbors plus the player itself, and there are two actions.
    }\label{tab:Qtable}
    \end{table}	

Strategy updating is driven by the following Q-learning algorithm. Each agent $i$ maintains an independent action-value matrix (i.e., Q-table $Q_i(s, a)$, see Table. \ref{tab:Qtable}). The state space $\mathbb{S}$ encapsulates the focal agent's perception of neighorhood. To construct a parsimonious representation, the state $s_i(t)$ is defined as the total number of cooperators within the inclusive neighborhood (the agent itself plus its four neighbors), yielding $s_i(t) = a_i(t) + \sum_{j \in \Omega_i} a_j(t)$ and $s_i(t) \in \{0, 1, \ldots, 5\}$.

The Q-values are updated synchronously according to the Bellman equation based on the effective fitness $F_i(t)$ of Eq. \eqref{eq:fitness}:
\begin{equation}
\begin{aligned}
			Q_i(s_i(t), a_i(t)) \leftarrow (1-\alpha) Q_i(s_i(t), a_i(t))\\ +
			 \alpha \Big[ F_i(t) + \gamma \max_{a'} Q_i(s_i(t+1), a') \Big],
\end{aligned}
\label{eq:bellman}
\end{equation}
where $\alpha \in (0,1]$ is the learning rate that controls the assimilation speed of new experiences, and $\gamma \in (0,1]$ is the discount factor that determines the weight of future returns. To balance exploration and exploitation, the agent adopts the $\varepsilon$-greedy policy: with probability $1 - \varepsilon$, it acts according to the guidance of the Q-table by selecting $\arg\max_{a} Q_i(s_i(t), a)$; with probability $\varepsilon$, it explores a random action.
	
Besides, agents undergo reputation updates based on their current actions. A cooperative act increases the reputation by a discrete step $c$, whereas defection diminishes it by $c$. To maintain rigorous mathematical boundaries, the reputation update rule is strictly defined as:
\begin{equation}
		R_i(t+1) = \max \Big( 0, \min \big( R_{max}, R_i(t) + c \cdot (2a_i(t) - 1) \big) \Big).
		\label{eq:update_repu}
\end{equation}
Since $a_i \in \{1, 0\}$, the term $(2a_i(t) - 1)$ elegantly outputs $+1$ for cooperation and $-1$ for defection. When $R_{max} = 0$, the reputation mechanism collapses, and the model is then structurally reduced to a standard spatial Q-learning PD game.

We implement a synchronous multi-agent simulation framework to execute the evolutionary dynamics. Initially, the elements of the Q-table of each agent are assigned random values drawn uniformly from $[0, 1]$. Initial actions are assigned randomly with equal probability, and initial reputations are uniformly distributed in $[0, R_{max}]$. Unless otherwise specified, the system size is set to $N = L \times L = 100 \times 100$. The hyperparameters of Q-learning are fixed at $\alpha = 0.1$, $\gamma = 0.9$, and $\varepsilon = 0.01$. Reputation step $c$ is set to $0.1$ with $R_{max} = 1.0$. The simulation ran a sufficiently long transient to achieve dynamic steady state, followed by an additional $1 \times 10 ^ 4$ steps at which macroscopic observables (such as ooperation prevalence $f_C $) are time averaged. The pseudo-code of our model is presented in Algorithm \ref{alg:rq}. All reported results are averaged over at least 20 independent realizations with different random initial conditions.

\begin{figure}[tbp]
		\centering
		\includegraphics[width=0.9\linewidth]{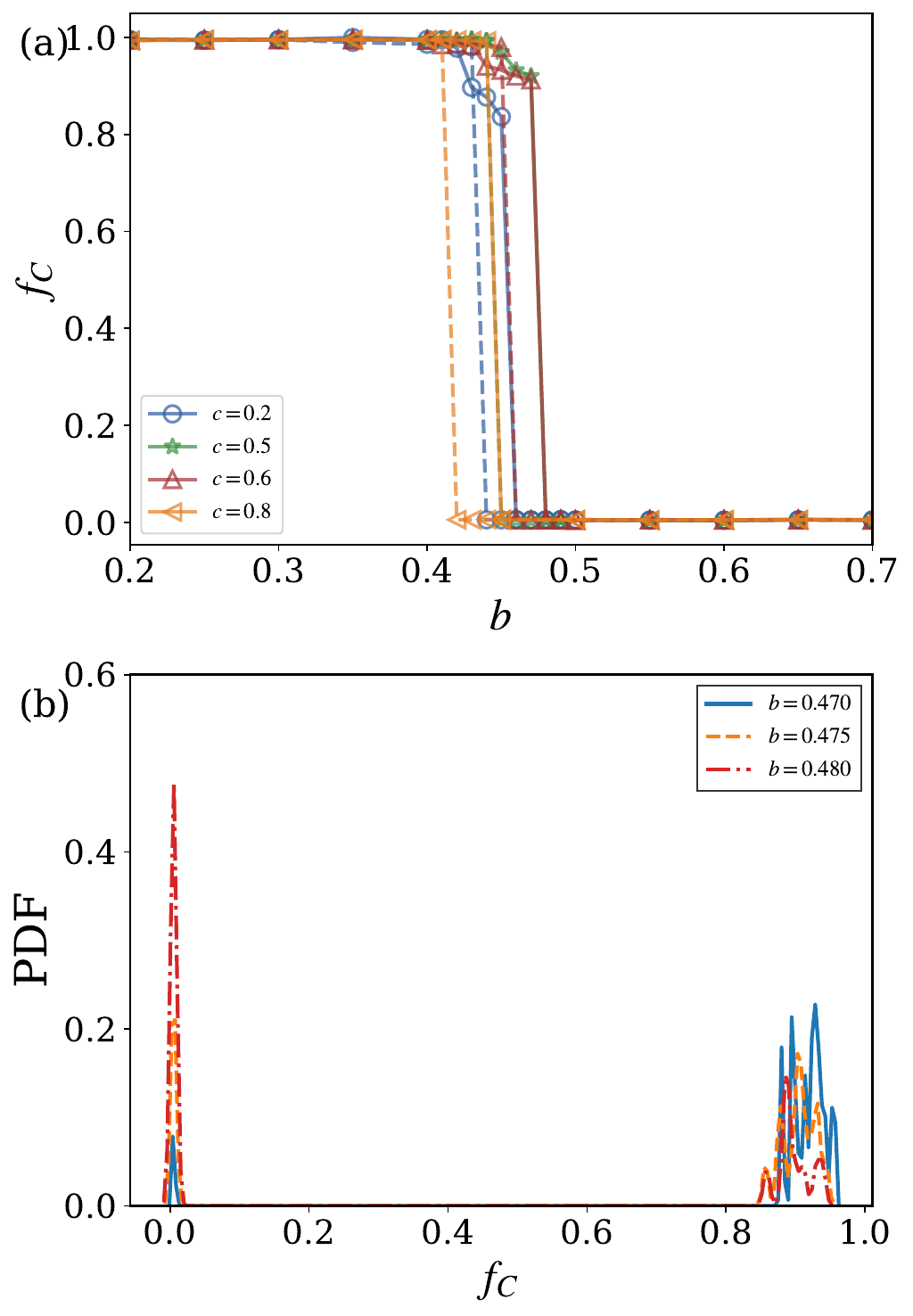}
		\caption{\textbf{Discontinuous phase transition of cooperation under reputation-modulated dynamics. }
			(a) The level of cooperation $f_C$ as a function of the temptation parameter $b$ for different reputation update rates $c$. The dashed region indicates the bistable regime obtained from different random initial configurations.
			(b) Probability density function (PDF) of the cooperation level $f_C$, revealing the coexistence between cooperative and defective states near the discontinuous transition. Other parameters: $c=0.5$ in panel (b).}
		\label{fig:double}
\end{figure}

	\begin{figure*}[htbp]
		\centering

		\includegraphics[width=0.94\linewidth]{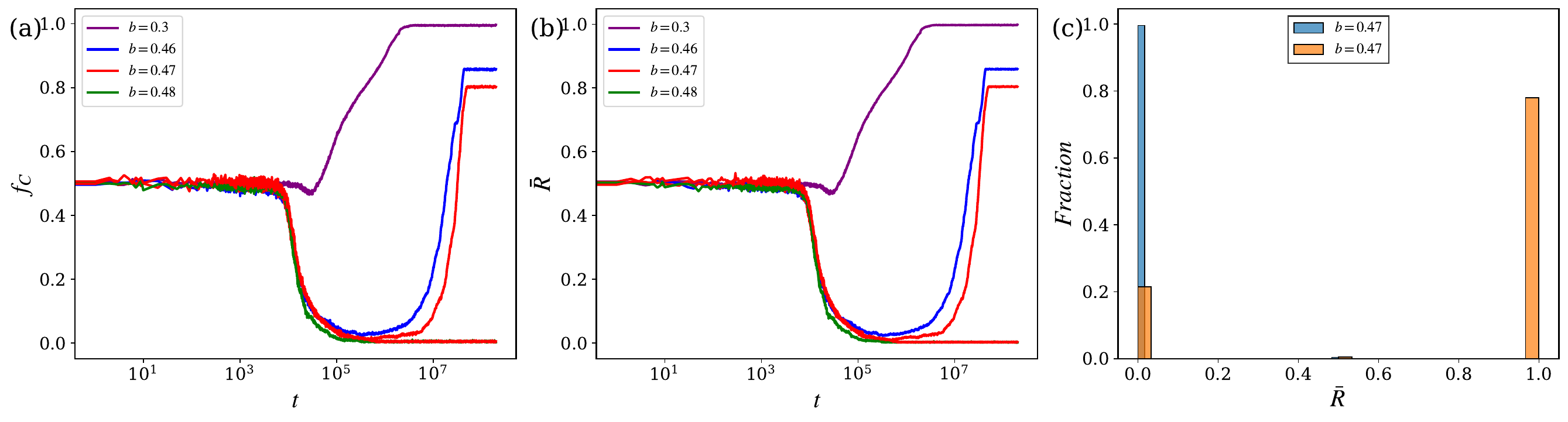}
		\caption{
		(a) Temporal evolution of the cooperation level $f_C$ for four chosen values of $b$ (shown in the legend). The red curve ($b=0.47$) exhibits bistable dynamics under random initial conditions, indicating the coexistence of cooperative and defective attractors near the critical regime. 
		(b) The corresponding temporal evolution of the average reputation of (a). 
		(c) Bimodal distribution of steady-state reputation for $b=0.47$ under random initial conditions, showing the spontaneous polarization of agents into low- and high-reputation groups.
        Other parameters: $c = 0.5$, $\alpha=0.1$, $\gamma=0.9$.}
        \label{fig:time_series}
	\end{figure*}

\section{Result}\label{sec3}
	
We first report the phase transition of cooperation prevalence $f_C$ for the population as a function of the temptation parameter $b$ for different reputation steps $c$, as shown in Fig.~\ref{fig:double}. Our results show that the introduction of reputation feedback significantly enhances cooperation at the population level. For all tested values of $c$, the system can evolve toward the fully cooperative state when the temptation $b\le 0.42$.
As $b$ increases beyond this range, the system undergoes an abrupt transition from full cooperation to full defection, suggesting the presence of a critical threshold. Through multiple independent implementations with different random initial configurations, we identified a bistable region that exhibits hysteresis like behavior, which is a characteristic of discontinuous (first-order) phase transitions.

This feature is further corroborated in Fig.~\ref{fig:double}(b), where we fix $c=0.5$ and present the steady-state cooperation level $f_C$ obtained from multiple runs with random initial conditions. The resulting bimodal distribution clearly reveals two competing attractors. As $b$ increases, the probability weight gradually shifts from the high-cooperation branch to the low-cooperation one, accompanied by a corresponding decrease in $f_C$.

To gain a deeper understanding of how reputation feedback shapes the evolutionary dynamics, we examine representative time series of the system. Fig.~\ref{fig:time_series}(a) shows the temporal evolution of $f_C$ for several values of $b$ at a fixed reputation update step $c=0.5$. In all cases, the dynamics first enter an RL transient exploration of approximately $10^4 $, during which the strategy is widely sampled and the level of cooperation fluctuates around $0.5$. However, the subsequent evolutionary trajectory strongly depends on the magnitude of temptation. 
For relatively small $b$, cooperation level rises rapidly after the exploration phase, and the system quickly converges to a highly cooperative state. In contrast, when 
$b$ approaches the critical threshold, the dynamics become markedly non-monotonic. Owing to the short-term payoff advantage of defection, the cooperation level drops sharply at first, yielding a near-fully defective metastable state. Nevertheless, due to the persistent exploration inherent in RL, alternative actions continue to be sampled and reinforced over time. After a sufficiently long period ($\sim10^{5}$ steps), the gradual accumulation of reputation amplifies cooperative tendencies, eventually allowing cooperation to dominate the population and stabilizing the system in a high-cooperation state.

	
We further illustrate the typical time series of bistable dynamics for $b=0.47$ in Fig.~\ref{fig:time_series} (a), where the outcomes of two evolutionary dynamics diverge dramatically, i.e., one leading to full defection ($f_C\rightarrow0$) and the other converging to a highly cooperative ($f_C\approx 0.8$). For larger values of $b$, the system irreversibly evolves toward full defection, without restoration of cooperation.

To complement the analysis of cooperation prevalence, we present the corresponding time series of the average reputation $\bar{R}$, shown in Fig.~\ref{fig:time_series}(b). At the beginning, each agent is assigned a random reputation value uniformly drawn from $[0,1]$, resulting in $\bar{R}(t=0)\approx0.5$. Given that reputation is directly tied to individual strategic behavior, its temporal evolution closely parallels that of the cooperation level and exhibits nearly identical dynamical features. During the early stage of the evolution with $f_C\approx 0.5$, the average reputation remains stable around $0.5$, as expected.

Over time, however, clusters of high-reputation agents gradually emerge (see Fig.~\ref{fig:snap}), enabling cooperators to resist invasion by defectors through spatial aggregation. As a consequence, the reputation of cooperative agents continues to rise, whereas that of defectors steadily declines. This picture is quantitatively confirmed in Fig.~\ref{fig:time_series}(c), which shows that the distribution of the steady-state of reputation for the representative bistable case $b=0.47$ exhibits a pronounced bimodal structure, clearly distinguishing persistent cooperators from persistent defectors. Interestingly, even within the bistable regime, a fraction of individuals continue to switch between cooperation and defection. Due to the repeated alternation between reputation-enhancing cooperative actions and reputation-decreasing defective behaviors, these individuals maintain intermediate reputation levels around $R\approx 0.5$.

\begin{figure*}[htbp]
		\centering
		\includegraphics[width=0.8\linewidth]{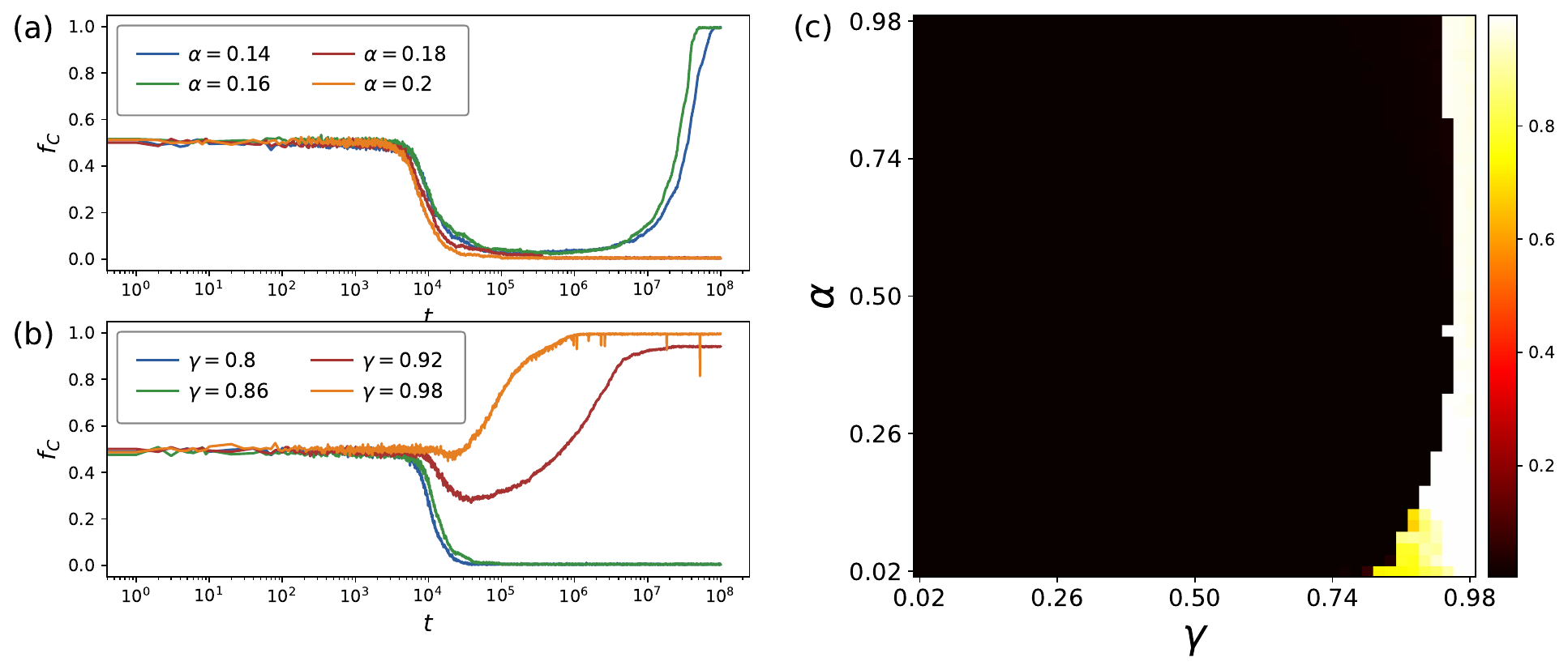}
		\caption{
		    (a) Time evolution of the cooperation level for different learning rates $\alpha$ at a fixed discount factor $\gamma=0.9$. 
			(b) Time evolution for different discount factors $\gamma$ at a fixed learning rate $\alpha=0.1$. 
			(c) Phase diagram in the $(\alpha,\gamma)$ parameter space. The color encodes the steady-state value of $f_C \in [0, 1]$. Note that the extreme cases $\alpha=0,1$ and $\gamma=0,1$ are excluded, as they correspond to trivial or degenerate learning dynamics. Accordingly, both parameters are restricted to the interval $[0.02,0.98]$.
			Other parameters are: $\epsilon=0.01$, $b=0.42$, and $c=0.2$. 
			}\label{fig:learning-parameters}
	\end{figure*}

Next, we examine the influence of two key parameters in Q-learning: the learning rate $\alpha$ and the discount factor $\gamma$. The learning rate $\alpha$ determines the speed at which agents adjust their Q-values in response to new experiences, whereas the discount factor $\gamma$ controls the balance between immediate and future rewards. Fig.~\ref{fig:learning-parameters} (a) demonstrates that, for a fixed discount factor $\gamma=0.9$, increasing $\alpha$ drives the system from a highly cooperative state toward full defection. This suggests that learning rate $\alpha$ tends to overwrite historical experience, thereby undermining the emergence of cooperation even in the presence of reputation feedback. In particular, agents converge prematurely to strategies that favor short-term payoff maximization, diminishing the long-term reinforcement effects facilitated by reputation accumulation.

In contrast, Fig.~\ref{fig:learning-parameters}(b) demonstrates that, for a fixed learning rate $\alpha$, increasing the discount factor $\gamma$ leads to a pronounced enhancement of cooperation. This behavior highlights the importance of long-term reward evaluation in cooperative evolution. A larger $\gamma$ enables agents to assign greater weight to future returns, allowing them to better recognize the delayed collective benefits generated by cooperative interactions. 

The complete phase diagram in the parameter space $(\alpha,\gamma)$ is presented in Fig.~\ref{fig:learning-parameters}(c), where cooperation is found to emerge only within the regime of small $\alpha$ and large $\gamma$. These findings highlight the fundamental interplay between historical experience and future-oriented decision-making: cooperation thrives when agents combine slow learning dynamics with a strong orientation towards future rewards~\cite{Ding2023Emergence,zheng2024evolution}.

	﻿
	﻿
	﻿

	﻿
	﻿

\section{Mechanism analysis}\label{sec4}

\begin{figure*}[htbp]
\centering
\includegraphics[width=0.8\linewidth]{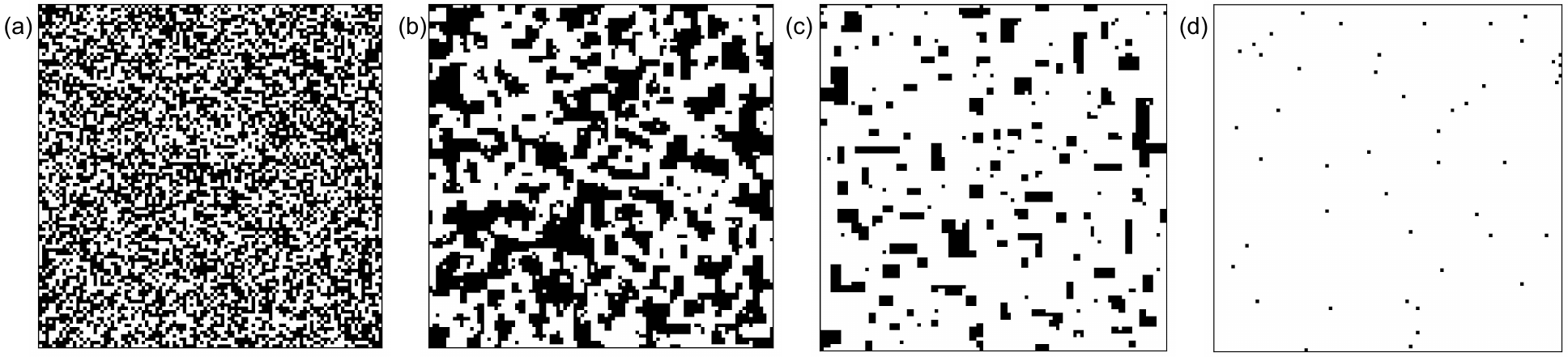}
\includegraphics[width=0.8\linewidth]{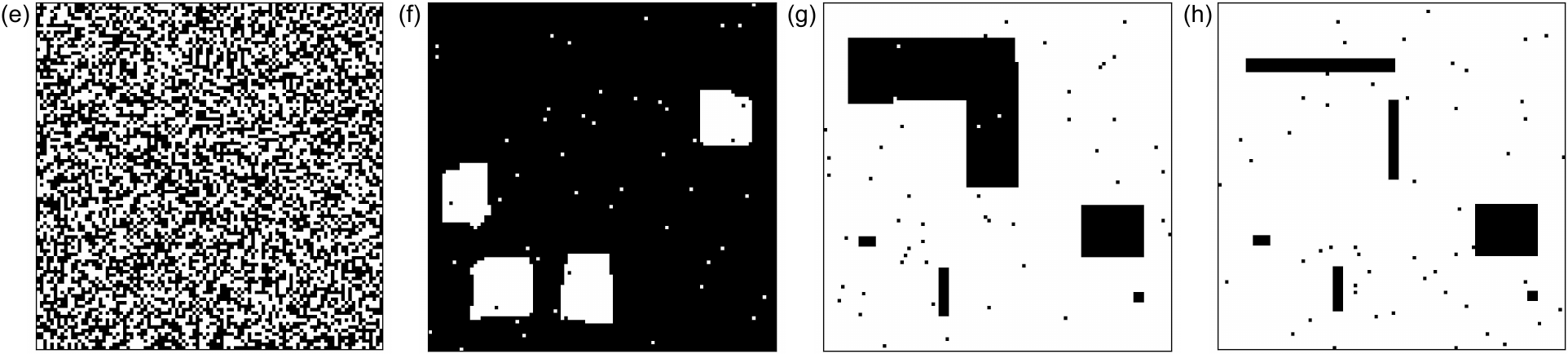}
\includegraphics[width=0.8\linewidth]{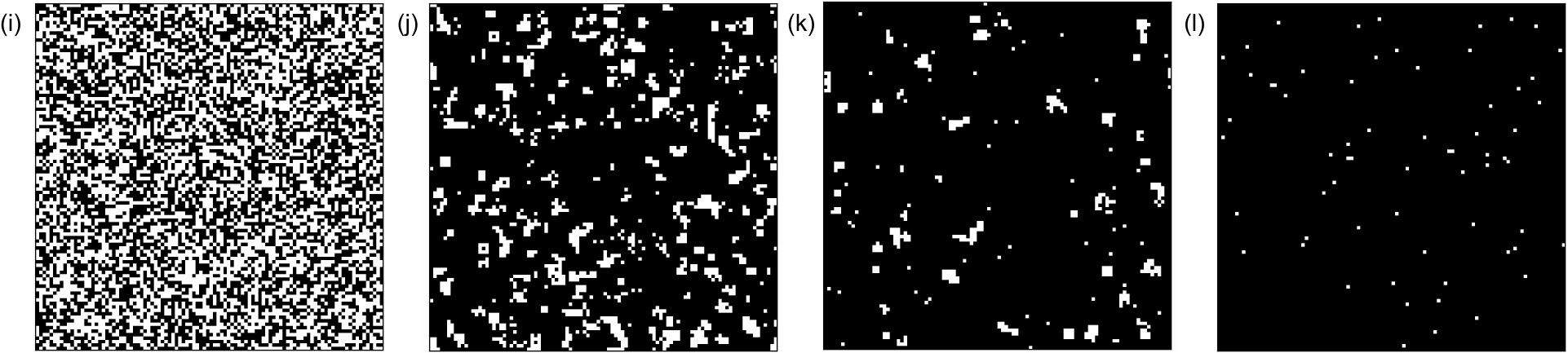}
\caption{\textbf{Typical spatial configurations illustrating the evolution of cooperation at a fixed reputation update rate $c=0.5$.} Cooperators and defectors are represented by white and black sites, respectively.
(a-d) Snapshots at $t=0$, $10^{5}$, $10^{6}$, and $10^{7}$ for $b=0.3$, showing the rapid expansion of cooperative domains toward a highly cooperative state.
(e-h) Snapshots at $t=0$, $10^{7}$, $7\times10^{7}$, and $15\times10^{7}$ for the left side of the transition point $b = 0.47$, where cooperation emerges through the nucleation and growth of cooperative clusters.
(i-l) Snapshots at $t=0$, $10^{4}$, $2\times10^{4}$, and $10^{5}$ for the defective side of the transition point $b = 0.47$, illustrating the gradual extinction of cooperative domains and convergence to the fully defective state. }
\label{fig:snap}
\end{figure*}

To understand the mechanism behind the emergence of cooperation, we present two representative spatial evolution snapshots under dilemma intensity in Figs. ~\ref {fig:snap}: weak dilemma region b=0.3 and bistable region near discontinuous transition $b=0.47$. Figs. ~\ref{fig:snap}(a-d) show that reputation feedback in the case of $b=0.3$ rapidly promotes cooperative behavior from the early stage of the evolution. Cooperative agents quickly become dominant and progressively eliminate surrounding defectors, ultimately driving the system toward the fully cooperative state.
	
By contrast, at the larger dilemma strength $b=0.47$, where the system enters the bistable regime, the evolutionary dynamics become strongly dependent on initial configurations. During the early evolution, defectors increase in density because of their short-term payoff advantage, as shown in Fig.~\ref{fig:snap}(f) and (j). However, when cooperative clusters survive and reach a sufficient size, reputation feedback enables these clusters to overcome local invasion by defectors. The surviving cooperative domains then undergo nucleation-driven growth, merging and expanding throughout the system until reaching a cooperative-dominated state. Notably, even in the long-term steady state, small defective clusters can still survive, which explains that the ultimate the prevalence of cooperation $f_C$ is smaller than 1(see Fig.~\ref{fig:time_series}(a)).

In addition, Figs.~\ref{fig:snap} (i-l) show that the nucleation-like growth process may also fail in this scenario, where the cooperative clusters shrink and vanish, and the system becomes fully defective.

	
	
To characterize how cooperation emerges at the individual level, we define the \emph{cooperation preference} as
	\begin{equation}
		\overline{\Delta Q}_{s_j}
		=
		\frac{1}{N}
		\sum_{i=1}^{N}
		\left(
		Q^{i}_{s_j,C}
		-
		Q^{i}_{s_j,D}
		\right),
		\label{eq:DeltaQ}
	\end{equation}
where $s_j \in \mathbb{S}$ denotes a specific state $j$ of the agent $i$ (Tab. \ref{tab:Qtable}). A positive value of $\overline{\Delta Q}_{s_j}$ indicates that agents in state $s_j$ prefer cooperation on average, whereas $\overline{\Delta Q}_{s_j}<0$ indicates a preference for defection.
	
\begin{figure}[htbp]
\centering
\includegraphics[width=0.95\linewidth]{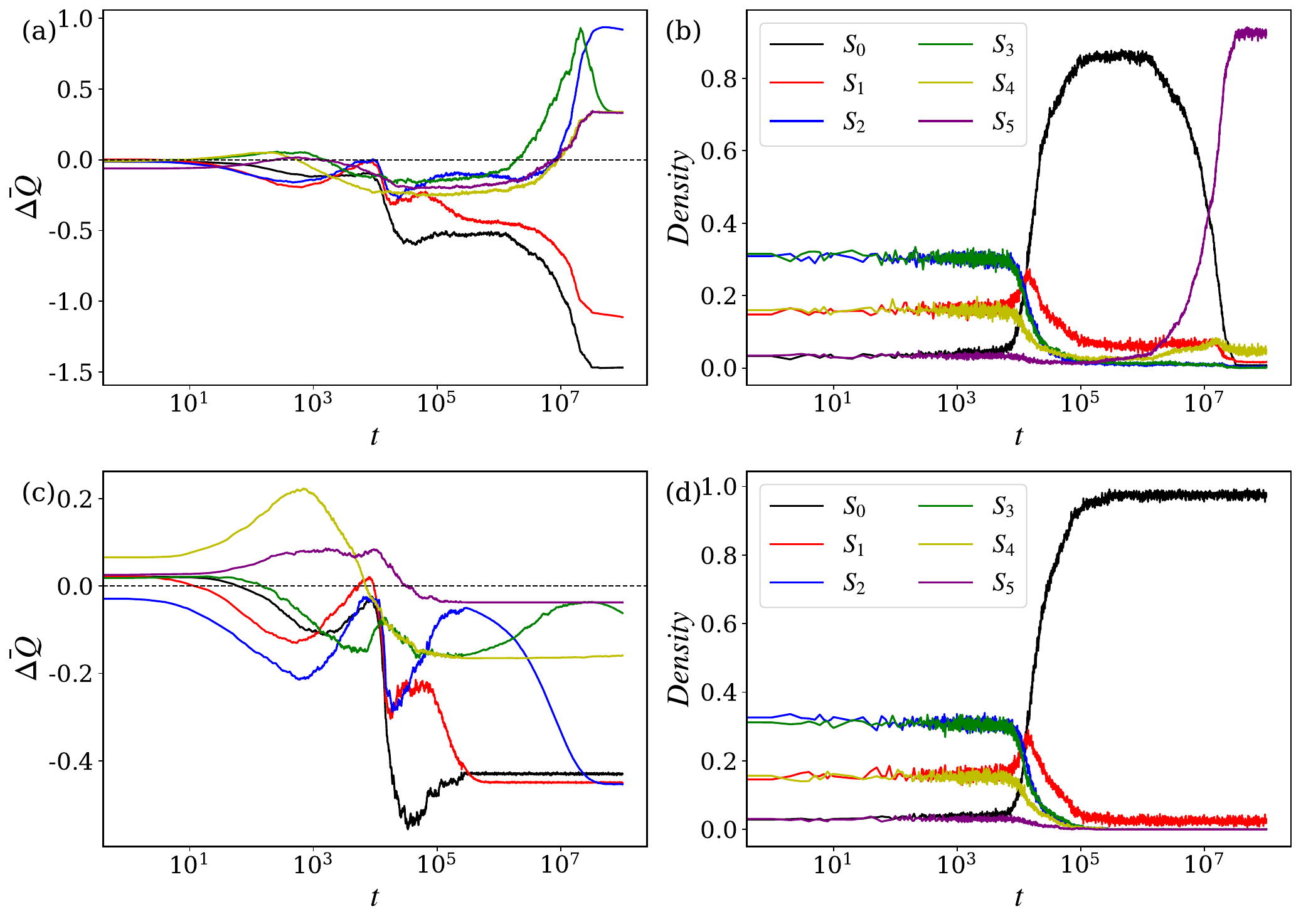}
\caption{
(a, c) Temporal evolution of the average cooperation preference $\overline{\Delta Q}_{s_j}$ for all states $s_j$ and (b, d) are the corresponding evolution of state densities when the system shows bistable states at $b = 0.45$. (a) in the high-cooperation and (c) the low-cooperation scenario.
In (b), the densities of states $s_{4,5}$ rise at $t\approx 10^5$, signaling cluster growth. In (d), all densities except for $s_0$ continue to decrease around $t\approx 10^4$, and the system collapses.
Other parameters: $c=0.5$, $\alpha=0.1$, $\gamma=0.9$, $\varepsilon=0.05$.
}
\label{fig:Q_table}
\end{figure}

We focus on the bistable regime where two representative stationary states are observed at $b=0.45$ and $c=0.5$, corresponding to the high-cooperation and low-cooperation scenarios, and examine the temporal evolution of the cooperation preference $\overline{\Delta Q}_{s_j}$ and the associated state densities, shown in Fig.~\ref{fig:Q_table}.

We find that agents with Q-learning develop fundamentally different Q-table structures in these two scenarios. For the high-cooperation scenario, agents learn to reciprocate cooperation with cooperation, namely $\overline{\Delta Q}_{s_{2,3,4,5}}>0$ (Fig.~\ref{fig:Q_table}(a)). This implies that once cooperative clusters emerge, they become self-reinforcing and expand rapidly, eventually eliminating surrounding defectors. By contrast, the agents in the log-cooperation scenario exhibit an entirely different preference. As shown in Fig.~\ref{fig:Q_table}(c), $\overline{\Delta Q}_{s_j}<0$ for all states, which means that cooperation fails to recover once agents are trapped in the defective neighborhood.
	
Figs.~\ref{fig:Q_table}(b,d) further demonstrate the corresponding evolution of the state densities. After the initial stage $t>10^4$, defectors quickly become dominant in both scenarios. However, the subsequent evolution diverges dramatically. In the high-cooperation scenario, after $t \approx 10^5$, the densities of states $s_{4,5}$ increase, indicating the growth of cooperative clusters [see Figs.~\ref{fig:snap}(e-h)]. As cooperative domains expand, defectors are gradually eliminated and the system ultimately converges to full cooperation, in which the density of $s_5$ dominates. In contrast, for the low-cooperation scenario, all densities except for $s_0$ vanish as no positive cooperation preference is observed, as shown in Fig.~\ref{fig:Q_table}(c). As a consequence, large cooperative clusters fail to emerge [see Figs.~\ref{fig:snap} (i-l)], and the system consequently evolves irreversibly toward the fully defective state.

	
	\begin{figure}[bpth]
		\centering
		\includegraphics[width=0.99\linewidth]{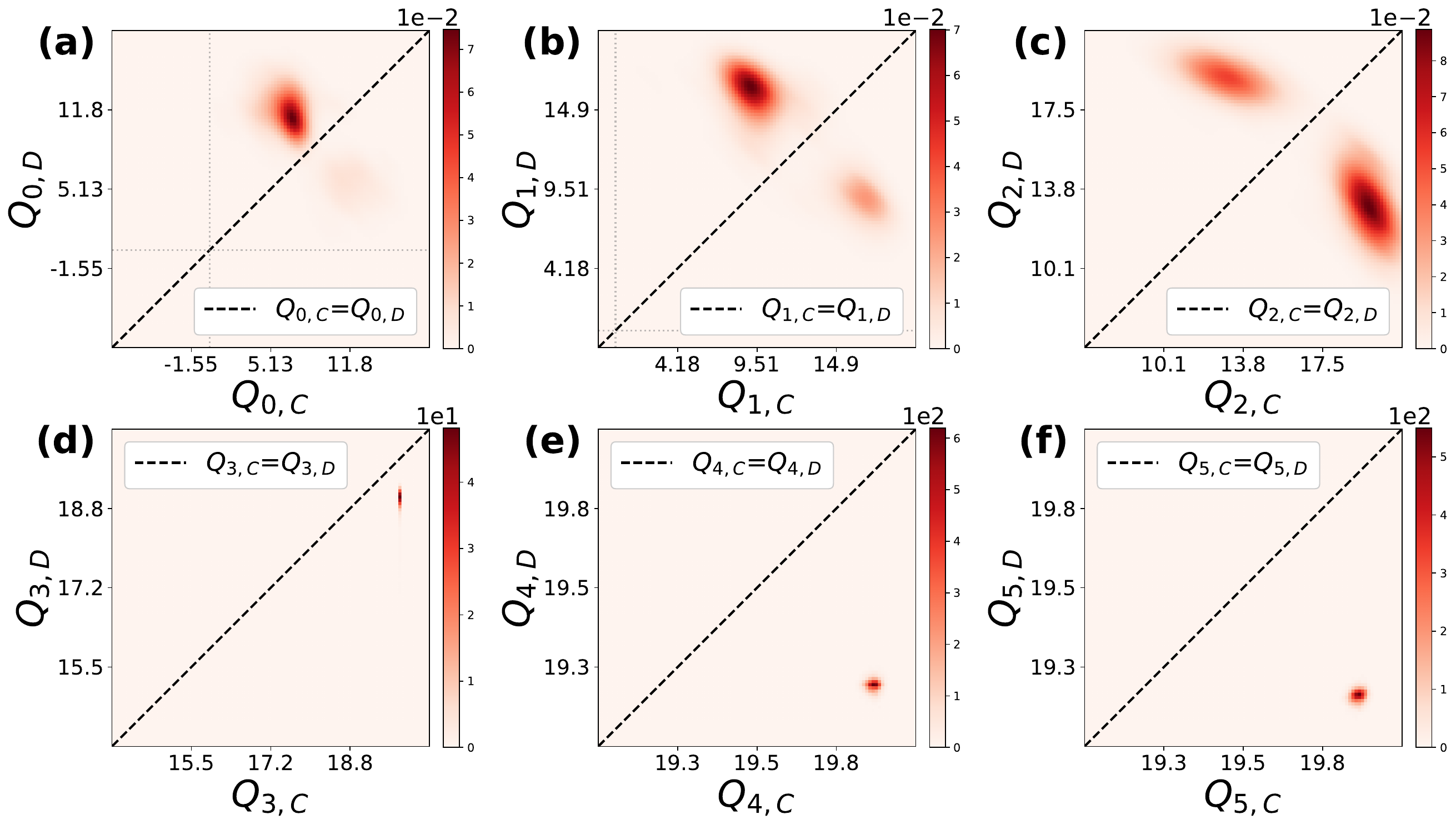}
		\includegraphics[width=0.99\linewidth]{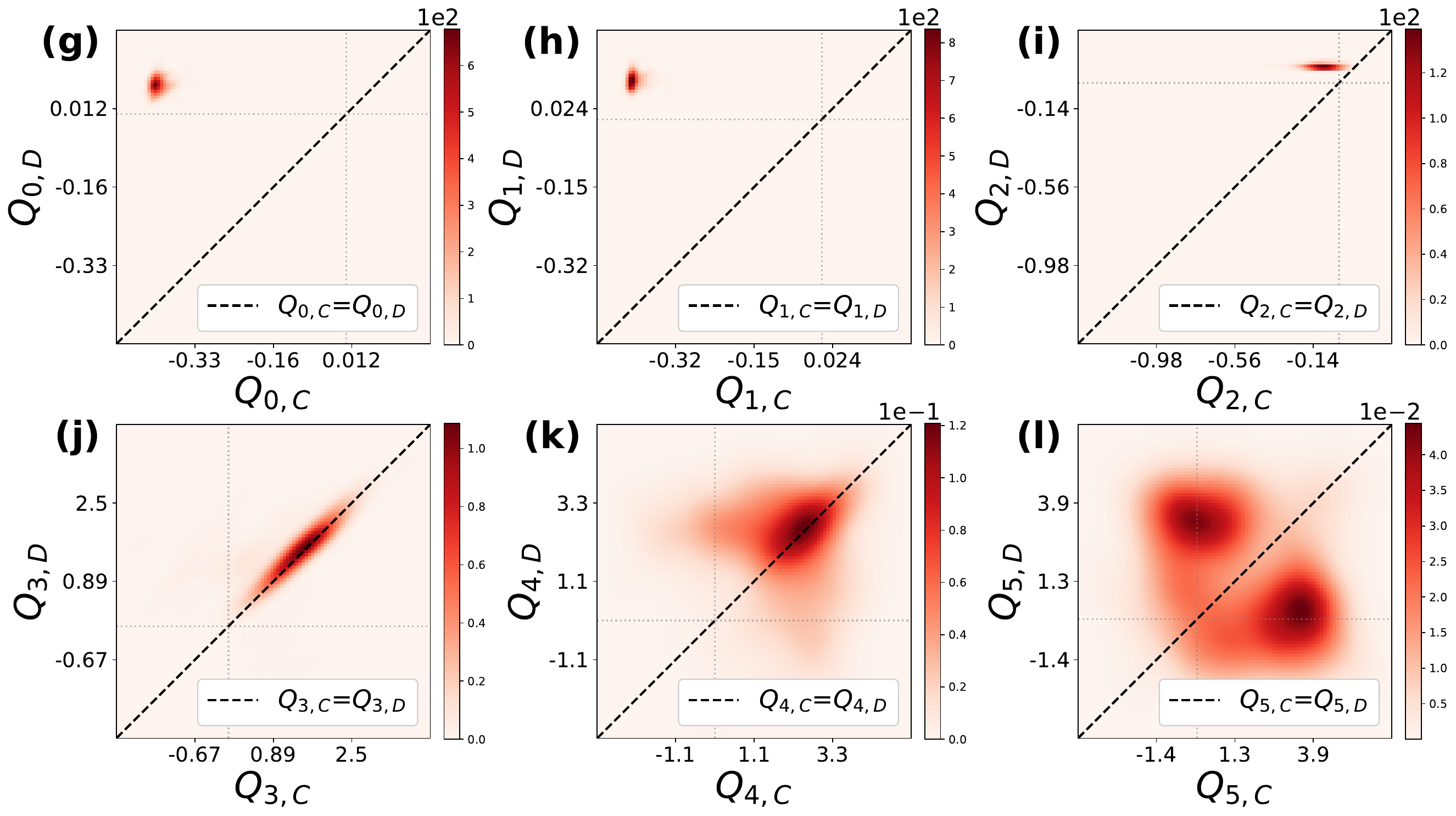}
		\caption{
		\textbf{\textbf{Probability density functions (PDFs) of state-action values in the bistable regime at $b=0.45$ .}} Each panel shows the distribution of Q values associated with a particular state $s_j$. The diagonal line $Q_{s_j,C}=Q_{s_j,D}$ separates cooperative and defective preferences: points below (above) the diagonal correspond to a preference for cooperation (defection). Panels (a-f) correspond to the high-cooperation branch, while panels (g-l) correspond to the low-cooperation branch. The value of colorbar represents the probability density per unit area in a two-dimensional Q-value space.}
		\label{fig:Q_value}
	\end{figure}
		
\begin{figure}[bpth]
\centering
\includegraphics[width=0.99\linewidth]{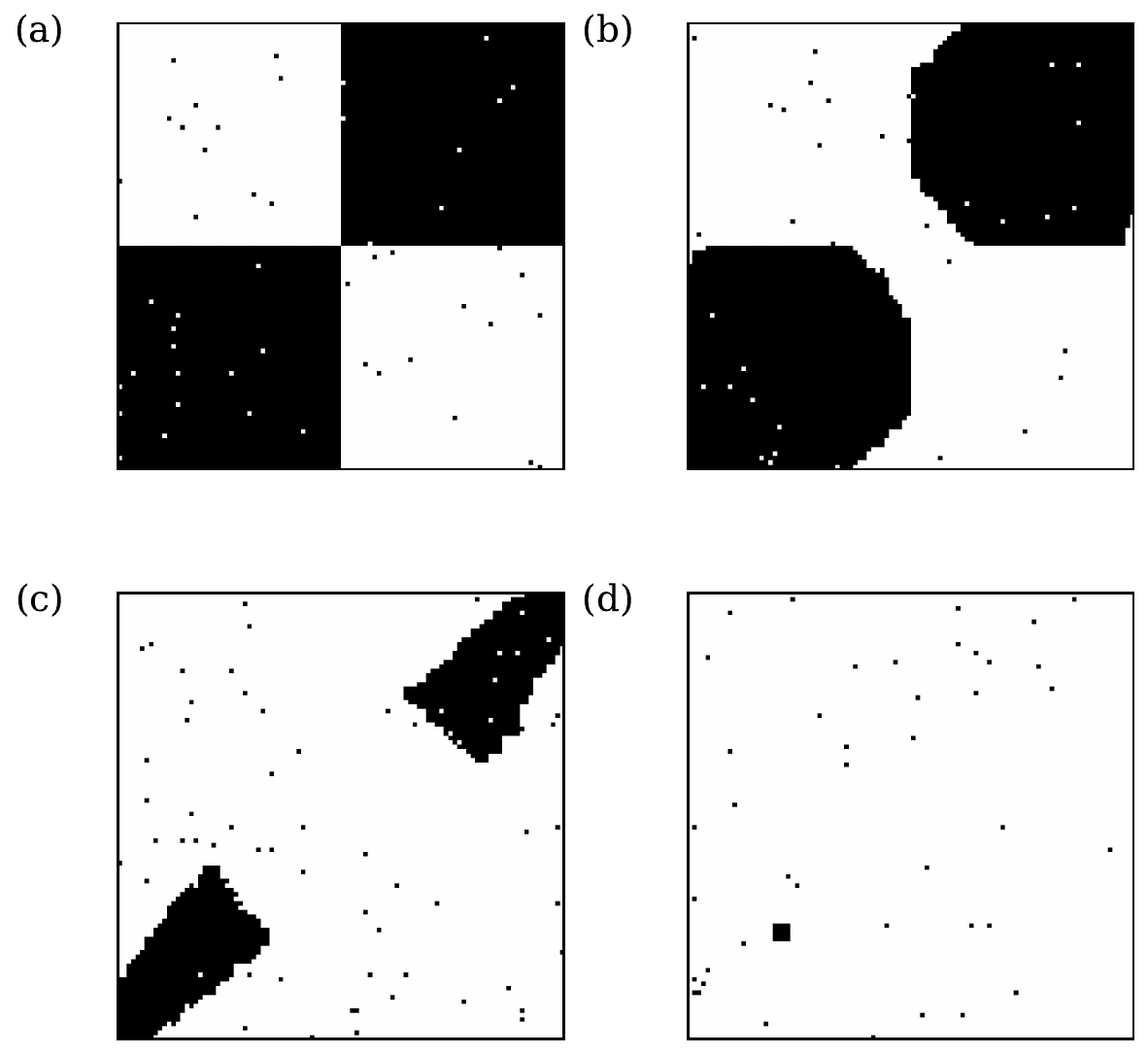}
\caption{\textbf{Evolution with a chosen initial state.}
(a) high-reputation and high-cooperation, (b) low-reputation and high-cooperation, (c) high-reputation and low-cooperation, and (d) low-reputation and low-cooperation initial conditions. Snapshots are taken at $t=0$, $3\times10^6$, $10^7$, $4\times10^7$ Monte Carlo steps, respectively for (a-d). 
Parameters: $c= 0.5$ , $p= 1$, $\rho=0.2$ and $b = 0.1$.}
\label{fig:snapshot_prepared}
\end{figure}
		
To corroborate the above analysis, we further present the probability density functions (PDF of the Q-values associated with different states in $b=0.45$, shown in Figs.~\ref{fig:Q_value}. In this figure, points above the diagonal ($Q_{s,C}=Q_{s,D}$) correspond to a preference for defection as $Q_{s,D}>Q_{s,C}$, whereas those below the diagonal indicate a preference for cooperation since $Q_{s,C}>Q_{s,D}$. Furthermore, deviations from the diagonal quantifies the robustness of the learned preference against stochastic exploration; the larger the deviations from the diagonal, the more robust the preference.


	
For the high-cooperation scenario [Figs.~\ref{fig:Q_value}(a-f)], the PDFs for states $s_{3,4,5}$ exhibit a single dominant peak in the cooperative region ($Q_{s,C}>Q_{s,D}$), indicating that agents unanimously learn to cooperate in the cooperative neighborhood. For less cooperative cases, $s_{0,1,2}$, the distributions split into two symmetric subpopulations around the diagonal. As the number of cooperative neighbors increases, the fraction of agents that favor cooperation increases.
As expected, PDFs of the low-cooperation scenario exhibit a qualitatively different distribution [Figs.~\ref{fig:Q_value}(g-l)]. While nearly all agents prefer defection within states $s_{0,1,2}$, the distributions in states $s_{3,4,5}$ remain approximately symmetric with respect to the diagonal. This occurs because these states are rarely visited during the evolutionary process, resulting in insufficient updates of the corresponding Q-values. Consequently, the initial randomness in these state-action values is largely preserved, preventing agents from developing a pronounced preference toward either cooperation or defection in these states.

	
The spatiotemporal evolution from the chosen initial conditions further clarifies how these learned microscopic rules generate macroscopic cooperation (see Fig.~\ref{fig:snapshot_prepared}). The left panels correspond to initially high-reputation populations, whereas the right panels correspond to initially low-reputation populations. 
As the system evolves, the increased interface roughness creates more opportunities for boundary defectors to encounter locally cooperative environments. In particular, whenever a defector acquires two or more cooperative neighbors, it enters states $s_{2,3,4,5}$. As demonstrated by the learned Q-table structure in Fig.~\ref{fig:Q_table}(a) and the Q-value distributions in Figs.~\ref{fig:Q_value}(a-f), agents in these states have a strong preference for cooperation. Consequently, boundary defectors are likely to cooperate, causing cooperative domains to expand. Cooperation therefore propagates through a nucleation-and-growth process driven by local reinforcement at interfaces. This microscopic mechanism provides a direct explanation for the emergence and persistence of large-scale cooperation under reputation-modulated learning.

\section{A pair-approximation theory}
To gain analytical insight into the reputation-modulated Q-learning dynamics, we develop a pair approximation that explicitly incorporates nearest-neighbor spatial correlations. Since the microscopic Q-tables retain the individuals' histories, an exact analytical treatment is generally intractable. Instead, we adopt an adiabatic approximation in which the Q-values are assumed to converge rapidly to their expected values given the current spatial configuration, allowing us to replace the instantaneous Q-value difference with the expected difference in the reputation-weighted reward signal. This approximation is justified by the small learning rate $\alpha = 0.1$, which implies that Q-values change slowly relative to the spatial configuration. Consequently, the spatial environment can be treated as quasi-stationary from the perspective of the Q-learning dynamics, enabling a closed-form analytical treatment of the spatial correlations. 
    
	
Let $p_{CC}$, $p_{CD}$, and $p_{DD}$ denote the densities of undirected nearest-neighbor pairs of types $CC$, $CD$, and $DD$, respectively. These quantities satisfy
\begin{equation}
	p_{CC}+p_{CD}+p_{DD}=1.
	\label{eq:pair_normalization}
\end{equation}
The corresponding densities of cooperators and defectors are
\begin{equation}
		\rho_C=p_{CC}+\frac{1}{2}p_{CD},
		\qquad
		\rho_D=p_{DD}+\frac{1}{2}p_{CD}=1-\rho_C.
		\label{eq:rho_from_pair}
\end{equation}
	
To characterize local environments, we focus on the conditional probability that a neighbor of a cooperator is cooperative,	
\begin{equation}
		q_{C|C} = \frac{2p_{CC}}{2p_{CC}+p_{CD}},
		\label{eq:qcc}
\end{equation}
and the conditional probability that a neighbor of a defector is cooperative,
\begin{equation}
		q_{C|D} = \frac{p_{CD}}{p_{CD}+2p_{DD}}.
		\label{eq:qcd}
\end{equation}
	
Following the standard pair-approximation closure, we assume that the states of the four neighbors are conditionally independent once the state of the focal player is specified. Therefore, the probability that a focal cooperator has $k$ cooperative neighbors is
\begin{equation}
		P(k|C) = \binom{4}{k} q_{C|C}^{k} (1-q_{C|C})^{4-k},
		\label{eq:pk_given_c}
\end{equation}
whereas for a focal defector,
\begin{equation}
		P(k|D) = \binom{4}{k} q_{C|D}^{k} (1-q_{C|D})^{4-k}.
		\label{eq:pk_given_d}
\end{equation}
	
The reputation-weighted reward depends on the average reputation of the local neighborhood. For a focal player surrounded by $k$ cooperators, we approximate the normalized neighborhood reputation as
\begin{equation}
		r(k) = \frac{k\mathcal R_C + (4-k)\mathcal R_D}{4R_{\max}},
		\label{eq:local_reputation_pair}
\end{equation}
where $\mathcal R_C$ and $\mathcal R_D$ denote the average reputations of cooperative and defective players, respectively.
	
According to the reward definition in Eq.~(\ref{eq:fitness}), cooperation sacrifices a direct payoff advantage $b$ while simultaneously generating a positive payoff externality for neighboring players. The effective reward difference between cooperation and defection is therefore approximated by
\begin{equation}
		\Delta F(k) = F_C(k)-F_D(k) = -[1-r(k)]b + r(k)\frac{1+b}{4}.
		\label{eq:fitness_difference_pair}
\end{equation}
The first term represents the immediate payoff disadvantage of cooperation, whereas the second term quantifies the reputation-weighted social benefit generated by cooperative behavior. Cooperation becomes locally favorable whenever $\Delta F(k)>0$, which yields the reputation threshold
\begin{equation}
		r(k) > r_c = \frac{4b}{1+5b}.
		\label{eq:reputation_threshold_pair}
\end{equation}
The above Eq.~(\ref{eq:reputation_threshold_pair}) highlights the role of reputation in the learning process. Although reputation does not alter the underlying game payoff matrix, it modifies the reward signal that guides reinforcement learning. Consequently, sufficiently reputable neighborhoods can transform the positive externality generated by cooperation into an effective learning advantage.
	
To approximate the $\varepsilon$-greedy Q-learning rule, we map the effective reward difference onto a probabilistic action-selection function. The probability of choosing cooperation is 
\begin{equation}
		\Phi_C(k) = \frac{\varepsilon}{2} + (1-\varepsilon) \frac{1}{1+\exp[-\beta\Delta F(k)]},
		\label{eq:soft_epsilon_pair}
\end{equation}
where $\varepsilon$ is the exploration rate and $\beta$ controls the sensitivity to reward differences. The probability of choosing defection is
\begin{equation}
		\Phi_D(k)=1-\Phi_C(k).
		\label{eq:phi_d}
\end{equation}
	
The hard $\varepsilon$-greedy rule is recovered in the limit $\beta\rightarrow\infty$. The average transition rates between the two strategies are
\begin{equation}
		W_{C\rightarrow D} = \sum_{k=0}^{4} P(k|C)\Phi_D(k),
		\label{eq:w_c_to_d}
\end{equation}
and	
\begin{equation}
		W_{D\rightarrow C} = \sum_{k=0}^{4} P(k|D)\Phi_C(k).
		\label{eq:w_d_to_c}
\end{equation}
	
To explicitly account for the local spatial correlations induced by the two-dimensional lattice, we track the dynamics of the three types of nearest-neighbor pairs, $CC$, $CD$, and $DD$. Consider a focal cooperator with $k$ cooperative neighbors and $4-k$ defective neighbors. When the focal player switches from cooperation to defection, each of its $k$ $CC$ links is converted into a $CD$ link, whereas each of its $(4-k) CD$ links is converted into a $DD$ link. Thus, a single $C\rightarrow D$ transition systematically rearranges the local pair structure: the number of $CC$ pairs decreases by $k$, the number of $DD$ pairs increases by $4-k$, while the number of $CD$ pairs changes accordingly,

\begin{subequations}
		\begin{align}
			\Delta p_{CC}^{C\rightarrow D}(k) &= -\frac{k}{4},\\
			\Delta p_{CD}^{C\rightarrow D}(k) &= \frac{2k-4}{4},\\
			\Delta p_{DD}^{C\rightarrow D}(k) &= \frac{4-k}{4}.
		\end{align}
		\label{eq:pair_change_c_to_d}
\end{subequations}
Similarly, the $D\rightarrow C$ transition produces
\begin{subequations}
		\begin{align}
			\Delta p_{CC}^{D\rightarrow C}(k) &= \frac{k}{4},\\
			\Delta p_{CD}^{D\rightarrow C}(k) &= \frac{4-2k}{4},\\
			\Delta p_{DD}^{D\rightarrow C}(k) &= -\frac{4-k}{4}.
		\end{align}
		\label{eq:pair_change_d_to_c}
\end{subequations}
	
Combining Eqs.~(\ref{eq:pk_given_c})-(\ref{eq:pair_change_d_to_c}), the pair-density dynamics becomes
\begin{subequations}
		\begin{align}
			\dot p_{CC} ={}& \rho_C \sum_{k=0}^{4} P(k|C)\Phi_D(k) \left( -\frac{k}{4} \right) \nonumber\\
			&+ \rho_D \sum_{k=0}^{4} P(k|D)\Phi_C(k) \left( \frac{k}{4} \right),\\
			\dot p_{CD} ={}& \rho_C \sum_{k=0}^{4} P(k|C)\Phi_D(k) \left( \frac{2k-4}{4} \right) \nonumber\\
			&+ \rho_D \sum_{k=0}^{4} P(k|D)\Phi_C(k) \left( \frac{4-2k}{4} \right),\\
			\dot p_{DD} ={}& \rho_C \sum_{k=0}^{4} P(k|C)\Phi_D(k) \left( \frac{4-k}{4} \right) \nonumber\\
			&+ \rho_D \sum_{k=0}^{4} P(k|D)\Phi_C(k) \left( -\frac{4-k}{4} \right).
		\end{align}
		\label{eq:pair_density_dynamics}
\end{subequations}
These equations preserve the normalization condition in Eq.~(\ref{eq:pair_normalization}) and describe the evolution of local spatial correlations.

To close the system, we introduce coarse-grained equations for the average reputations of cooperators and defectors. Consistent with the microscopic updating rule, cooperators gradually accumulate reputation whereas defectors gradually lose reputation. In the continuous-time limit, the reputation dynamics are approximated by
\begin{subequations}
		\begin{align}
			\frac{d\mathcal R_C}{dt} &= c \left(1-\frac{\mathcal R_C}{R_{\max}}\right),\\
			\frac{d\mathcal R_D}{dt} &= -c \frac{\mathcal R_D}{R_{\max}}.
		\end{align}
		\label{eq:dynamic_reputation_closure}
\end{subequations}
The stationary solution of Eqs.~(\ref{eq:dynamic_reputation_closure}) is
\begin{equation}
		\mathcal R_C = R_{\max},\qquad \mathcal R_D = 0.
		\label{eq:stationary_reputation_closure}
\end{equation}

\begin{figure} 
		\centering \includegraphics[width=0.8\linewidth]{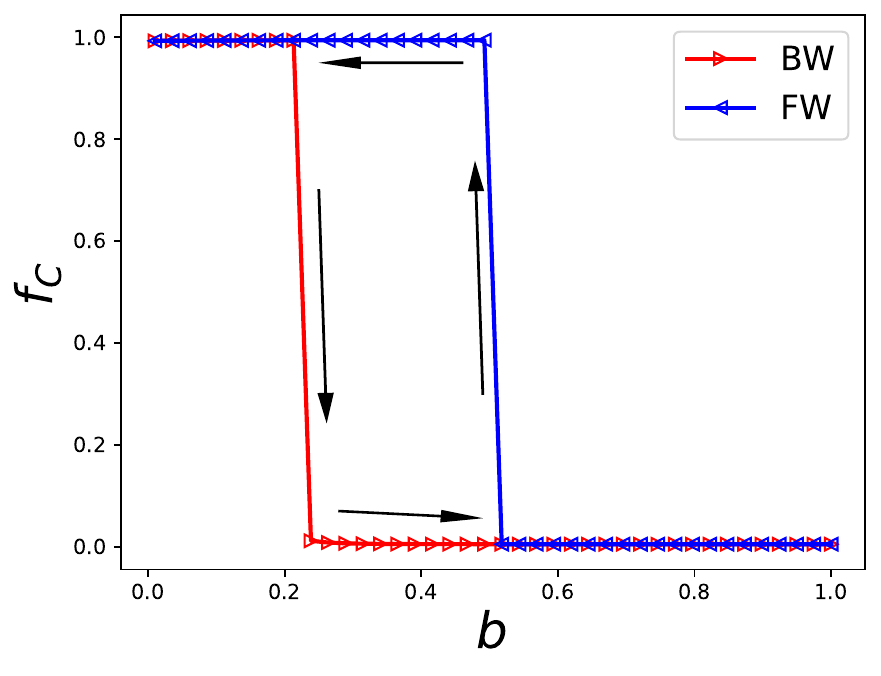} 
		\caption{The theoretical prediction (Eq. \ref{eq:pair_density_dynamics}) of steady-state cooperation prevalence $f_C$ as a function of the dilemma strength $b$. The blue and red curves correspond to forward (FW) and backward (BW) processes, respectively. The resulting hysteresis loop indicates the presence of bistability and a first-order phase transition. Parameters are set as $\varepsilon=0.01, c=0.5, \beta=25.0$.} 
		\label{fig:theory} 
\end{figure}

Although our above theoretical framework neglects higher-order spatial correlations and the heterogeneous distribution of Q-values, the adiabatic approximation combined with different initial conditions successfully reproduces the discontinuous transition between cooperative and defective phases and captures the emergence of a hysteretic regime (see Fig.~\ref{fig:theory}). 
In addition, the forward bifurcation analysis accurately predicts the transition threshold observed in simulations ($b \approx 0.45$) and qualitatively recovers the cooperative enhancement induced by reputation feedback. These results demonstrate that the proposed approximation, despite its reduced microscopic complexity, preserves the essential macroscopic phase behavior and provides analytical insights into how reputation-driven feedback promotes the emergence and stability of cooperation.

\section{Conclusion and discussion}\label{sec:conclusion}
	


Motivated by experimental evidence that individuals adjust their weighting of social information based on its perceived reliability~\cite{Lamba2014Social,molleman2014consistent}, we have investigated a reputation-modulated reinforcement learning framework within a spatial prisoner's dilemma game. In contrast to conventional reputation models, reputation in our setup neither alters the payoff matrix nor modifies the interaction network. Instead, it governs how agents integrate personal and social information during the value updating process. Our results demonstrate that reputation-modulated learning substantially enhances cooperation in a broad range of dilemma strengths. More importantly, increasing the temptation to defect induces a discontinuous transition from a highly cooperative state to a defection-dominated one. Near the transition point, the system exhibits pronounced bistability, where cooperative and defective outcomes coexist under identical parameter configurations. Spatiotemporal analyses further reveal that this transition is driven by the nucleation of cooperative clusters, and in the bistable regime, the formation of such clusters critically determines the eventual fate of the system. 
	

	﻿

The underlying mechanism of these macroscopic phenomena can be understood through the evolutionary dynamics of Q-tables. As learning proceeds, agents gradually develop state-dependent behavioral preferences: local cooperative environments reinforce cooperative actions, whereas unfavorable environments favor defection~\cite{shengCatalytic2024,ZhaoEvolution2025}. Consequently, reputation establishes a positive feedback loop—cooperation enhances reputation, reputation amplifies the influence of socially acquired information, and the resulting learning dynamics further stabilize cooperative behavior. A pair-approximation analysis offers a simple theoretical interpretation of this process. Although material rewards consistently favor defection, cooperative actions simultaneously generate positive externalities for neighboring agents. When local reputation exceeds the critical threshold $r_c = 4b/(1+5b)$ (Eq. \eqref{eq:reputation_threshold_pair}, i.e., socially acquired information becomes sufficiently influential to offset the short-term advantage of defection, making cooperation locally favorable. This threshold underscores the essential role of spatial clustering and reputation polarization: cooperation does not arise as isolated altruism, but rather as a collective state that fundamentally reshapes the effective learning environment.

Recent studies have highlighted the crucial role of reputation in promoting cooperation from different perspectives. For example, reputation-based incentive mechanisms have been shown to facilitate cooperation by modifying evaluation processes and enhancing the fitness advantage of cooperative individuals~\cite{Hu2024Reputation}. Other studies have incorporated reputation into imitation dynamics, demonstrating that reputation-dependent strategy updating can stabilize cooperation even under strong social dilemmas~\cite{Feng2024An}. Beyond individual-level interactions, reputation has also been coupled with adaptive network evolution, where individuals selectively maintain connections with high-reputation neighbors, thereby reshaping the underlying interaction structure and promoting cooperative clusters~\cite{zhang2024Reputation}. 

More broadly, our work demonstrates that reputation can promote cooperation without directly incentivizing cooperative behavior or penalizing defection. By modulating how adaptive agents process and utilize social information, reputation alone proves capable of generating stable cooperation, bistability, and abrupt collective transitions. These findings offer a fresh perspective on the interplay among reputation, reinforcement learning, and collective behavior, suggesting that information-mediated learning, where reputation signals the reliability of social information, may serve as a general and parsimonious mechanism for sustaining cooperation across a wide range of complex adaptive systems, without requiring external incentives or institutional enforcement.

	
	﻿
	
	﻿
	
	﻿
	
	﻿
	

	\section*{Code availability}
	The code for generating key results in this study is available at \href{https://github.com/chenli-lab/RL-CC}{https://github.com/chenli-lab/RL-Reputation}.

	\section*{Acknowledgments}
    This work is supported by the National Natural Science Foundation of China (Grants Nos. 12075144, 12165014 and 42461144209), the Fundamental Research Funds for the Central Universities (Grant No. GK202401002), the Natural Science Basic Research Program of Shaanxi (Grants No. 2026JC-YBMS-0012), the Shanghai Natural Science Foundation (Grant No. 24ZR1420700), and the STI2030-Major Projects (grant no. 2021ZD0202600).
	
\bibliography{References}

\begin{thebibliography}{58}%
\makeatletter
\providecommand \@ifxundefined [1]{%
 \@ifx{#1\undefined}
}%
\providecommand \@ifnum [1]{%
 \ifnum #1\expandafter \@firstoftwo
 \else \expandafter \@secondoftwo
 \fi
}%
\providecommand \@ifx [1]{%
 \ifx #1\expandafter \@firstoftwo
 \else \expandafter \@secondoftwo
 \fi
}%
\providecommand \natexlab [1]{#1}%
\providecommand \enquote  [1]{``#1''}%
\providecommand \bibnamefont  [1]{#1}%
\providecommand \bibfnamefont [1]{#1}%
\providecommand \citenamefont [1]{#1}%
\providecommand \href@noop [0]{\@secondoftwo}%
\providecommand \href [0]{\begingroup \@sanitize@url \@href}%
\providecommand \@href[1]{\@@startlink{#1}\@@href}%
\providecommand \@@href[1]{\endgroup#1\@@endlink}%
\providecommand \@sanitize@url [0]{\catcode `\\12\catcode `\$12\catcode
  `\&12\catcode `\#12\catcode `\^12\catcode `\_12\catcode `\%12\relax}%
\providecommand \@@startlink[1]{}%
\providecommand \@@endlink[0]{}%
\providecommand \url  [0]{\begingroup\@sanitize@url \@url }%
\providecommand \@url [1]{\endgroup\@href {#1}{\urlprefix }}%
\providecommand \urlprefix  [0]{URL }%
\providecommand \Eprint [0]{\href }%
\providecommand \doibase [0]{http://dx.doi.org/}%
\providecommand \selectlanguage [0]{\@gobble}%
\providecommand \bibinfo  [0]{\@secondoftwo}%
\providecommand \bibfield  [0]{\@secondoftwo}%
\providecommand \translation [1]{[#1]}%
\providecommand \BibitemOpen [0]{}%
\providecommand \bibitemStop [0]{}%
\providecommand \bibitemNoStop [0]{.\EOS\space}%
\providecommand \EOS [0]{\spacefactor3000\relax}%
\providecommand \BibitemShut  [1]{\csname bibitem#1\endcsname}%
\let\auto@bib@innerbib\@empty
\bibitem [{\citenamefont {Nielsen}(1985)}]{Nielsen1985cooperation}%
  \BibitemOpen
  \bibfield  {author} {\bibinfo {author} {\bibfnamefont {R.~P.}\ \bibnamefont
  {Nielsen}},\ }\href {\doibase 10.2307/257983} {\bibfield  {journal} {\bibinfo
   {journal} {The Academy of Management Review}\ }\textbf {\bibinfo {volume}
  {10}},\ \bibinfo {pages} {368} (\bibinfo {year} {1985})}\BibitemShut
  {NoStop}%
\bibitem [{\citenamefont {Colman}(1995)}]{Colman1995Theory}%
  \BibitemOpen
  \bibfield  {author} {\bibinfo {author} {\bibfnamefont {A.}~\bibnamefont
  {Colman}},\ }\href {\doibase 10.4324/9780203761335} {\emph {\bibinfo {title}
  {Game Theory and Its Applications in the Social and Biological Sciences}}}\
  (\bibinfo  {publisher} {Psychology Press},\ \bibinfo {year}
  {1995})\BibitemShut {NoStop}%
\bibitem [{\citenamefont {Nowak}(2006)}]{Nowak2006Evolution}%
  \BibitemOpen
  \bibfield  {author} {\bibinfo {author} {\bibfnamefont {M.~A.}\ \bibnamefont
  {Nowak}},\ }\href {https://www.hup.harvard.edu/books/9780674023383} {\emph
  {\bibinfo {title} {Evolutionary Dynamics}}}\ (\bibinfo  {publisher}
  {Belknap/Harvard},\ \bibinfo {year} {2006})\BibitemShut {NoStop}%
\bibitem [{\citenamefont {Dawkins}(2006)}]{Dawkins2006selfish}%
  \BibitemOpen
  \bibfield  {author} {\bibinfo {author} {\bibfnamefont {R.}~\bibnamefont
  {Dawkins}},\ }\href {\doibase 10.1007/978-3-319-16999-6_1876-1} {\emph
  {\bibinfo {title} {The Selfish Gene}}}\ (\bibinfo  {publisher} {Oxford
  University Press},\ \bibinfo {address} {New York, US},\ \bibinfo {year}
  {2006})\BibitemShut {NoStop}%
\bibitem [{\citenamefont {Wilson}()}]{Wilson1975Sociobiology}%
  \BibitemOpen
  \bibfield  {author} {\bibinfo {author} {\bibfnamefont {E.~O.}\ \bibnamefont
  {Wilson}},\ }\href@noop {} {\emph {\bibinfo {title} {Sociobiology: The New
  Synthesis}}}\ (\bibinfo  {publisher} {Belknap Press of Harvard U Press},\
  \bibinfo {address} {Oxford, England})\BibitemShut {NoStop}%
\bibitem [{\citenamefont {Maynard~Smith}(1964)}]{Smith1964Group}%
  \BibitemOpen
  \bibfield  {author} {\bibinfo {author} {\bibfnamefont {J.}~\bibnamefont
  {Maynard~Smith}},\ }\href {\doibase 10.1038/2011145a0} {\bibfield  {journal}
  {\bibinfo  {journal} {Nature}\ }\textbf {\bibinfo {volume} {201}},\ \bibinfo
  {pages} {1145} (\bibinfo {year} {1964})}\BibitemShut {NoStop}%
\bibitem [{\citenamefont {Charlesworth}(2000)}]{Charlesworth2000Levels}%
  \BibitemOpen
  \bibfield  {author} {\bibinfo {author} {\bibfnamefont {B.}~\bibnamefont
  {Charlesworth}},\ }\href {\doibase 10.1046/j.1365-2540.2000.0726a.x}
  {\bibfield  {journal} {\bibinfo  {journal} {Heredity}\ }\textbf {\bibinfo
  {volume} {84}},\ \bibinfo {pages} {493} (\bibinfo {year} {2000})}\BibitemShut
  {NoStop}%
\bibitem [{\citenamefont {Pacheco}\ \emph {et~al.}(2008)\citenamefont
  {Pacheco}, \citenamefont {Traulsen}, \citenamefont {Ohtsuki},\ and\
  \citenamefont {Nowak}}]{Nowak2008Repeated}%
  \BibitemOpen
  \bibfield  {author} {\bibinfo {author} {\bibfnamefont {J.~M.}\ \bibnamefont
  {Pacheco}}, \bibinfo {author} {\bibfnamefont {A.}~\bibnamefont {Traulsen}},
  \bibinfo {author} {\bibfnamefont {H.}~\bibnamefont {Ohtsuki}}, \ and\
  \bibinfo {author} {\bibfnamefont {M.~A.}\ \bibnamefont {Nowak}},\ }\href
  {\doibase https://doi.org/10.1016/j.jtbi.2007.10.040} {\bibfield  {journal}
  {\bibinfo  {journal} {Journal of Theoretical Biology}\ }\textbf {\bibinfo
  {volume} {250}},\ \bibinfo {pages} {723} (\bibinfo {year}
  {2008})}\BibitemShut {NoStop}%
\bibitem [{\citenamefont {Wang}\ \emph
  {et~al.}(2013{\natexlab{a}})\citenamefont {Wang}, \citenamefont {Szolnoki},\
  and\ \citenamefont {Perc}}]{Perc2013Interdependent}%
  \BibitemOpen
  \bibfield  {author} {\bibinfo {author} {\bibfnamefont {Z.}~\bibnamefont
  {Wang}}, \bibinfo {author} {\bibfnamefont {A.}~\bibnamefont {Szolnoki}}, \
  and\ \bibinfo {author} {\bibfnamefont {M.}~\bibnamefont {Perc}},\ }\href
  {\doibase 10.1038/srep01183} {\bibfield  {journal} {\bibinfo  {journal}
  {Scientific Reports}\ }\textbf {\bibinfo {volume} {3}},\ \bibinfo {pages}
  {1183} (\bibinfo {year} {2013}{\natexlab{a}})}\BibitemShut {NoStop}%
\bibitem [{\citenamefont {Nowak}\ and\ \citenamefont
  {Sigmund}(1998{\natexlab{a}})}]{Nowak1998indirect}%
  \BibitemOpen
  \bibfield  {author} {\bibinfo {author} {\bibfnamefont {M.~A.}\ \bibnamefont
  {Nowak}}\ and\ \bibinfo {author} {\bibfnamefont {K.}~\bibnamefont
  {Sigmund}},\ }\href {\doibase 10.1038/31225} {\bibfield  {journal} {\bibinfo
  {journal} {Nature}\ }\textbf {\bibinfo {volume} {393}},\ \bibinfo {pages}
  {573} (\bibinfo {year} {1998}{\natexlab{a}})}\BibitemShut {NoStop}%
\bibitem [{\citenamefont {Ohtsuki}\ and\ \citenamefont
  {Iwasa}(2006{\natexlab{a}})}]{Ohtsuki2006indirect}%
  \BibitemOpen
  \bibfield  {author} {\bibinfo {author} {\bibfnamefont {H.}~\bibnamefont
  {Ohtsuki}}\ and\ \bibinfo {author} {\bibfnamefont {Y.}~\bibnamefont
  {Iwasa}},\ }\href {\doibase 10.1016/j.jtbi.2005.08.008} {\bibfield  {journal}
  {\bibinfo  {journal} {Journal of Theoretical Biology}\ }\textbf {\bibinfo
  {volume} {239}},\ \bibinfo {pages} {435} (\bibinfo {year}
  {2006}{\natexlab{a}})}\BibitemShut {NoStop}%
\bibitem [{\citenamefont {Nowak}\ and\ \citenamefont
  {May}(1992)}]{Nowak1992spatial}%
  \BibitemOpen
  \bibfield  {author} {\bibinfo {author} {\bibfnamefont {M.~A.}\ \bibnamefont
  {Nowak}}\ and\ \bibinfo {author} {\bibfnamefont {R.~M.}\ \bibnamefont
  {May}},\ }\href {\doibase 10.1038/359826a0} {\bibfield  {journal} {\bibinfo
  {journal} {Nature}\ }\textbf {\bibinfo {volume} {359}},\ \bibinfo {pages}
  {826} (\bibinfo {year} {1992})}\BibitemShut {NoStop}%
\bibitem [{\citenamefont {Szab{\'o}}\ and\ \citenamefont
  {T{\"{o}}ke}(1998)}]{Szabo1998Evolutionary}%
  \BibitemOpen
  \bibfield  {author} {\bibinfo {author} {\bibfnamefont {G.}~\bibnamefont
  {Szab{\'o}}}\ and\ \bibinfo {author} {\bibfnamefont {C.}~\bibnamefont
  {T{\"{o}}ke}},\ }\href {\doibase 10.1103/PhysRevE.58.69} {\bibfield
  {journal} {\bibinfo  {journal} {Physical Review E}\ }\textbf {\bibinfo
  {volume} {58}},\ \bibinfo {pages} {69} (\bibinfo {year} {1998})}\BibitemShut
  {NoStop}%
\bibitem [{\citenamefont {Wang}\ \emph
  {et~al.}(2013{\natexlab{b}})\citenamefont {Wang}, \citenamefont {Szolnoki},\
  and\ \citenamefont {Perc}}]{Wang2013Interdependent}%
  \BibitemOpen
  \bibfield  {author} {\bibinfo {author} {\bibfnamefont {Z.}~\bibnamefont
  {Wang}}, \bibinfo {author} {\bibfnamefont {A.}~\bibnamefont {Szolnoki}}, \
  and\ \bibinfo {author} {\bibfnamefont {M.}~\bibnamefont {Perc}},\ }\href
  {\doibase 10.1038/srep01183} {\bibfield  {journal} {\bibinfo  {journal}
  {Scientific Reports}\ }\textbf {\bibinfo {volume} {3}},\ \bibinfo {pages}
  {1183} (\bibinfo {year} {2013}{\natexlab{b}})}\BibitemShut {NoStop}%
\bibitem [{\citenamefont {Szolnoki}\ \emph {et~al.}(2010)\citenamefont
  {Szolnoki}, \citenamefont {Wang}, \citenamefont {Wang},\ and\ \citenamefont
  {Zhu}}]{Szolnoki2010Dynamically}%
  \BibitemOpen
  \bibfield  {author} {\bibinfo {author} {\bibfnamefont {A.}~\bibnamefont
  {Szolnoki}}, \bibinfo {author} {\bibfnamefont {Z.}~\bibnamefont {Wang}},
  \bibinfo {author} {\bibfnamefont {J.}~\bibnamefont {Wang}}, \ and\ \bibinfo
  {author} {\bibfnamefont {X.}~\bibnamefont {Zhu}},\ }\href {\doibase
  10.1103/PhysRevE.82.036110} {\bibfield  {journal} {\bibinfo  {journal}
  {Physical Review E}\ }\textbf {\bibinfo {volume} {82}},\ \bibinfo {pages}
  {036110} (\bibinfo {year} {2010})}\BibitemShut {NoStop}%
\bibitem [{\citenamefont {Liang}\ \emph {et~al.}(2022)\citenamefont {Liang},
  \citenamefont {Wang}, \citenamefont {Zhang}, \citenamefont {Zheng},
  \citenamefont {Ma},\ and\ \citenamefont {Chen}}]{liang2022dynamical}%
  \BibitemOpen
  \bibfield  {author} {\bibinfo {author} {\bibfnamefont {R.}~\bibnamefont
  {Liang}}, \bibinfo {author} {\bibfnamefont {Q.}~\bibnamefont {Wang}},
  \bibinfo {author} {\bibfnamefont {J.}~\bibnamefont {Zhang}}, \bibinfo
  {author} {\bibfnamefont {G.}~\bibnamefont {Zheng}}, \bibinfo {author}
  {\bibfnamefont {L.}~\bibnamefont {Ma}}, \ and\ \bibinfo {author}
  {\bibfnamefont {L.}~\bibnamefont {Chen}},\ }\href {\doibase
  https://doi.org/10.1103/PhysRevE.105.054302} {\bibfield  {journal} {\bibinfo
  {journal} {Physical Review E}\ }\textbf {\bibinfo {volume} {105}},\ \bibinfo
  {pages} {054302} (\bibinfo {year} {2022})}\BibitemShut {NoStop}%
\bibitem [{\citenamefont {Xia}\ \emph {et~al.}(2023)\citenamefont {Xia},
  \citenamefont {Wang}, \citenamefont {Perc},\ and\ \citenamefont
  {Wang}}]{xia2023reputation}%
  \BibitemOpen
  \bibfield  {author} {\bibinfo {author} {\bibfnamefont {C.}~\bibnamefont
  {Xia}}, \bibinfo {author} {\bibfnamefont {J.}~\bibnamefont {Wang}}, \bibinfo
  {author} {\bibfnamefont {M.}~\bibnamefont {Perc}}, \ and\ \bibinfo {author}
  {\bibfnamefont {Z.}~\bibnamefont {Wang}},\ }\href {\doibase
  https://doi.org/10.1016/j.plrev.2023.05.002} {\bibfield  {journal} {\bibinfo
  {journal} {Physics of life reviews}\ }\textbf {\bibinfo {volume} {46}},\
  \bibinfo {pages} {8} (\bibinfo {year} {2023})}\BibitemShut {NoStop}%
\bibitem [{\citenamefont {Fehr}\ and\ \citenamefont
  {Gächter}(2002)}]{Fehr2002punishment}%
  \BibitemOpen
  \bibfield  {author} {\bibinfo {author} {\bibfnamefont {E.}~\bibnamefont
  {Fehr}}\ and\ \bibinfo {author} {\bibfnamefont {S.}~\bibnamefont
  {Gächter}},\ }\href {\doibase 10.1038/415137a} {\bibfield  {journal}
  {\bibinfo  {journal} {Nature}\ }\textbf {\bibinfo {volume} {415}},\ \bibinfo
  {pages} {137} (\bibinfo {year} {2002})}\BibitemShut {NoStop}%
\bibitem [{\citenamefont {Herrmann}\ \emph {et~al.}(2008)\citenamefont
  {Herrmann}, \citenamefont {Th\"{o}ni},\ and\ \citenamefont
  {G{\"a}chter}}]{Herrmann2008Antisocial}%
  \BibitemOpen
  \bibfield  {author} {\bibinfo {author} {\bibfnamefont {B.}~\bibnamefont
  {Herrmann}}, \bibinfo {author} {\bibfnamefont {C.}~\bibnamefont {Th\"{o}ni}},
  \ and\ \bibinfo {author} {\bibfnamefont {S.}~\bibnamefont {G{\"a}chter}},\
  }\href {\doibase 10.1126/science.1153808} {\bibfield  {journal} {\bibinfo
  {journal} {Science}\ }\textbf {\bibinfo {volume} {319}},\ \bibinfo {pages}
  {1362} (\bibinfo {year} {2008})}\BibitemShut {NoStop}%
\bibitem [{\citenamefont {Clutton-Brock}\ and\ \citenamefont
  {Parker}(1995)}]{Clutton1995Punishment}%
  \BibitemOpen
  \bibfield  {author} {\bibinfo {author} {\bibfnamefont {T.~H.}\ \bibnamefont
  {Clutton-Brock}}\ and\ \bibinfo {author} {\bibfnamefont {G.~A.}\ \bibnamefont
  {Parker}},\ }\href {\doibase https://doi.org/10.1038/373209a0} {\bibfield
  {journal} {\bibinfo  {journal} {Nature}\ }\textbf {\bibinfo {volume} {373}},\
  \bibinfo {pages} {209–216} (\bibinfo {year} {1995})}\BibitemShut {NoStop}%
\bibitem [{\citenamefont {Diekmann}\ and\ \citenamefont
  {Przepiorka}(2015)}]{Diekmann2015Punitive}%
  \BibitemOpen
  \bibfield  {author} {\bibinfo {author} {\bibfnamefont {A.}~\bibnamefont
  {Diekmann}}\ and\ \bibinfo {author} {\bibfnamefont {W.}~\bibnamefont
  {Przepiorka}},\ }\href {\doibase 10.1038/srep10321} {\bibfield  {journal}
  {\bibinfo  {journal} {Scientific Reports}\ }\textbf {\bibinfo {volume} {5}},\
  \bibinfo {pages} {10321} (\bibinfo {year} {2015})}\BibitemShut {NoStop}%
\bibitem [{\citenamefont {Tilman}\ \emph {et~al.}(2020)\citenamefont {Tilman},
  \citenamefont {Plotkin},\ and\ \citenamefont
  {Ak{\c{c}}ay}}]{tilman2020evolutionary}%
  \BibitemOpen
  \bibfield  {author} {\bibinfo {author} {\bibfnamefont {A.~R.}\ \bibnamefont
  {Tilman}}, \bibinfo {author} {\bibfnamefont {J.~B.}\ \bibnamefont {Plotkin}},
  \ and\ \bibinfo {author} {\bibfnamefont {E.}~\bibnamefont {Ak{\c{c}}ay}},\
  }\href {\doibase https://doi.org/10.1038/s41467-020-14531-6} {\bibfield
  {journal} {\bibinfo  {journal} {Nature communications}\ }\textbf {\bibinfo
  {volume} {11}},\ \bibinfo {pages} {915} (\bibinfo {year} {2020})}\BibitemShut
  {NoStop}%
\bibitem [{\citenamefont {Weitz}\ \emph {et~al.}(2016)\citenamefont {Weitz},
  \citenamefont {Eksin}, \citenamefont {Paarporn}, \citenamefont {Brown},\ and\
  \citenamefont {Ratcliff}}]{Joshua2016An}%
  \BibitemOpen
  \bibfield  {author} {\bibinfo {author} {\bibfnamefont {J.~S.}\ \bibnamefont
  {Weitz}}, \bibinfo {author} {\bibfnamefont {C.}~\bibnamefont {Eksin}},
  \bibinfo {author} {\bibfnamefont {K.}~\bibnamefont {Paarporn}}, \bibinfo
  {author} {\bibfnamefont {S.~P.}\ \bibnamefont {Brown}}, \ and\ \bibinfo
  {author} {\bibfnamefont {W.~C.}\ \bibnamefont {Ratcliff}},\ }\href {\doibase
  10.1073/pnas.1604096113} {\bibfield  {journal} {\bibinfo  {journal}
  {Proceedings of the National Academy of Sciences}\ }\textbf {\bibinfo
  {volume} {113}},\ \bibinfo {pages} {E7518} (\bibinfo {year}
  {2016})}\BibitemShut {NoStop}%
\bibitem [{\citenamefont {Milinski}\ \emph {et~al.}(2002)\citenamefont
  {Milinski}, \citenamefont {Semmann},\ and\ \citenamefont
  {Krambeck}}]{milinski2002reputation}%
  \BibitemOpen
  \bibfield  {author} {\bibinfo {author} {\bibfnamefont {M.}~\bibnamefont
  {Milinski}}, \bibinfo {author} {\bibfnamefont {D.}~\bibnamefont {Semmann}}, \
  and\ \bibinfo {author} {\bibfnamefont {H.-J.}\ \bibnamefont {Krambeck}},\
  }\href {\doibase https://doi.org/10.1038/415424a} {\bibfield  {journal}
  {\bibinfo  {journal} {Nature}\ }\textbf {\bibinfo {volume} {415}},\ \bibinfo
  {pages} {424} (\bibinfo {year} {2002})}\BibitemShut {NoStop}%
\bibitem [{\citenamefont {Nowak}\ and\ \citenamefont
  {Sigmund}(1998{\natexlab{b}})}]{nowak1998evolution}%
  \BibitemOpen
  \bibfield  {author} {\bibinfo {author} {\bibfnamefont {M.~A.}\ \bibnamefont
  {Nowak}}\ and\ \bibinfo {author} {\bibfnamefont {K.}~\bibnamefont
  {Sigmund}},\ }\href {\doibase https://doi.org/10.1038/31225} {\bibfield
  {journal} {\bibinfo  {journal} {Nature}\ }\textbf {\bibinfo {volume} {393}},\
  \bibinfo {pages} {573} (\bibinfo {year} {1998}{\natexlab{b}})}\BibitemShut
  {NoStop}%
\bibitem [{\citenamefont {Ohtsuki}\ and\ \citenamefont
  {Iwasa}(2006{\natexlab{b}})}]{Ohtsuki2006The}%
  \BibitemOpen
  \bibfield  {author} {\bibinfo {author} {\bibfnamefont {H.}~\bibnamefont
  {Ohtsuki}}\ and\ \bibinfo {author} {\bibfnamefont {Y.}~\bibnamefont
  {Iwasa}},\ }\href {\doibase https://doi.org/10.1016/j.jtbi.2005.08.008}
  {\bibfield  {journal} {\bibinfo  {journal} {Journal of Theoretical Biology}\
  }\textbf {\bibinfo {volume} {239}},\ \bibinfo {pages} {435} (\bibinfo {year}
  {2006}{\natexlab{b}})}\BibitemShut {NoStop}%
\bibitem [{\citenamefont {Panchanathan}\ and\ \citenamefont
  {Boyd}(2004)}]{panchanathan2004indirect}%
  \BibitemOpen
  \bibfield  {author} {\bibinfo {author} {\bibfnamefont {K.}~\bibnamefont
  {Panchanathan}}\ and\ \bibinfo {author} {\bibfnamefont {R.}~\bibnamefont
  {Boyd}},\ }\href {\doibase https://doi.org/10.1038/nature02978} {\bibfield
  {journal} {\bibinfo  {journal} {Nature}\ }\textbf {\bibinfo {volume} {432}},\
  \bibinfo {pages} {499} (\bibinfo {year} {2004})}\BibitemShut {NoStop}%
\bibitem [{\citenamefont {Murase}\ and\ \citenamefont
  {Hilbe}(2023)}]{murase2023indirect}%
  \BibitemOpen
  \bibfield  {author} {\bibinfo {author} {\bibfnamefont {Y.}~\bibnamefont
  {Murase}}\ and\ \bibinfo {author} {\bibfnamefont {C.}~\bibnamefont {Hilbe}},\
  }\href {\doibase https://doi.org/10.1371/journal.pcbi.1011271} {\bibfield
  {journal} {\bibinfo  {journal} {PLOS Computational Biology}\ }\textbf
  {\bibinfo {volume} {19}},\ \bibinfo {pages} {e1011271} (\bibinfo {year}
  {2023})}\BibitemShut {NoStop}%
\bibitem [{\citenamefont {Liu}\ and\ \citenamefont
  {Chen}(2017)}]{Liu2017Sustainable}%
  \BibitemOpen
  \bibfield  {author} {\bibinfo {author} {\bibfnamefont {Y.}~\bibnamefont
  {Liu}}\ and\ \bibinfo {author} {\bibfnamefont {T.}~\bibnamefont {Chen}},\
  }\href {\doibase https://doi.org/10.1016/j.biosystems.2017.08.003} {\bibfield
   {journal} {\bibinfo  {journal} {Biosystems}\ }\textbf {\bibinfo {volume}
  {160}},\ \bibinfo {pages} {33} (\bibinfo {year} {2017})}\BibitemShut
  {NoStop}%
\bibitem [{\citenamefont {Dong}\ \emph {et~al.}(2019)\citenamefont {Dong},
  \citenamefont {Hao}, \citenamefont {Wang}, \citenamefont {Liu},\ and\
  \citenamefont {Xia}}]{DONG2019Cooperation}%
  \BibitemOpen
  \bibfield  {author} {\bibinfo {author} {\bibfnamefont {Y.}~\bibnamefont
  {Dong}}, \bibinfo {author} {\bibfnamefont {G.}~\bibnamefont {Hao}}, \bibinfo
  {author} {\bibfnamefont {J.}~\bibnamefont {Wang}}, \bibinfo {author}
  {\bibfnamefont {C.}~\bibnamefont {Liu}}, \ and\ \bibinfo {author}
  {\bibfnamefont {C.}~\bibnamefont {Xia}},\ }\href {\doibase
  https://doi.org/10.1016/j.physleta.2019.01.021} {\bibfield  {journal}
  {\bibinfo  {journal} {Physics Letters A}\ }\textbf {\bibinfo {volume}
  {383}},\ \bibinfo {pages} {1157} (\bibinfo {year} {2019})}\BibitemShut
  {NoStop}%
\bibitem [{\citenamefont {Han}\ \emph {et~al.}(2022)\citenamefont {Han},
  \citenamefont {Zhang}, \citenamefont {Sun},\ and\ \citenamefont
  {Xia}}]{HAN2022Role}%
  \BibitemOpen
  \bibfield  {author} {\bibinfo {author} {\bibfnamefont {W.}~\bibnamefont
  {Han}}, \bibinfo {author} {\bibfnamefont {Z.}~\bibnamefont {Zhang}}, \bibinfo
  {author} {\bibfnamefont {J.}~\bibnamefont {Sun}}, \ and\ \bibinfo {author}
  {\bibfnamefont {C.}~\bibnamefont {Xia}},\ }\href {\doibase
  https://doi.org/10.1016/j.chaos.2022.112385} {\bibfield  {journal} {\bibinfo
  {journal} {Chaos, Solitons \& Fractals}\ }\textbf {\bibinfo {volume} {161}},\
  \bibinfo {pages} {112385} (\bibinfo {year} {2022})}\BibitemShut {NoStop}%
\bibitem [{\citenamefont {Yang}\ \emph {et~al.}(2019)\citenamefont {Yang},
  \citenamefont {Wang},\ and\ \citenamefont {Xia}}]{YANG2019Evolution}%
  \BibitemOpen
  \bibfield  {author} {\bibinfo {author} {\bibfnamefont {W.}~\bibnamefont
  {Yang}}, \bibinfo {author} {\bibfnamefont {J.}~\bibnamefont {Wang}}, \ and\
  \bibinfo {author} {\bibfnamefont {C.}~\bibnamefont {Xia}},\ }\href {\doibase
  https://doi.org/10.1016/j.physleta.2019.07.014} {\bibfield  {journal}
  {\bibinfo  {journal} {Physics Letters A}\ }\textbf {\bibinfo {volume}
  {383}},\ \bibinfo {pages} {125826} (\bibinfo {year} {2019})}\BibitemShut
  {NoStop}%
\bibitem [{\citenamefont {Schmid}\ \emph {et~al.}(2023)\citenamefont {Schmid},
  \citenamefont {Ekbatani}, \citenamefont {Hilbe},\ and\ \citenamefont
  {Chatterjee}}]{schmid2023quantitative}%
  \BibitemOpen
  \bibfield  {author} {\bibinfo {author} {\bibfnamefont {L.}~\bibnamefont
  {Schmid}}, \bibinfo {author} {\bibfnamefont {F.}~\bibnamefont {Ekbatani}},
  \bibinfo {author} {\bibfnamefont {C.}~\bibnamefont {Hilbe}}, \ and\ \bibinfo
  {author} {\bibfnamefont {K.}~\bibnamefont {Chatterjee}},\ }\href {\doibase
  https://doi.org/10.1038/s41467-023-37817-x} {\bibfield  {journal} {\bibinfo
  {journal} {Nature Communications}\ }\textbf {\bibinfo {volume} {14}},\
  \bibinfo {pages} {2086} (\bibinfo {year} {2023})}\BibitemShut {NoStop}%
\bibitem [{\citenamefont {Zheng}\ \emph {et~al.}(2026)\citenamefont {Zheng},
  \citenamefont {Ou}, \citenamefont {Deng}, \citenamefont {Zhang},\ and\
  \citenamefont {Chen}}]{zheng2026brief}%
  \BibitemOpen
  \bibfield  {author} {\bibinfo {author} {\bibfnamefont {G.}~\bibnamefont
  {Zheng}}, \bibinfo {author} {\bibfnamefont {X.}~\bibnamefont {Ou}}, \bibinfo
  {author} {\bibfnamefont {S.}~\bibnamefont {Deng}}, \bibinfo {author}
  {\bibfnamefont {J.}~\bibnamefont {Zhang}}, \ and\ \bibinfo {author}
  {\bibfnamefont {L.}~\bibnamefont {Chen}},\ }\href {\doibase
  https://doi.org/10.1088/1572-9494/ae503e} {\bibfield  {journal} {\bibinfo
  {journal} {Communications in Theoretical Physics}\ }\textbf {\bibinfo
  {volume} {78}},\ \bibinfo {pages} {067601} (\bibinfo {year}
  {2026})}\BibitemShut {NoStop}%
\bibitem [{\citenamefont {Tomov}\ \emph {et~al.}(2021)\citenamefont {Tomov},
  \citenamefont {Schulz},\ and\ \citenamefont {Gershman}}]{TomovMulti2021}%
  \BibitemOpen
  \bibfield  {author} {\bibinfo {author} {\bibfnamefont {M.~S.}\ \bibnamefont
  {Tomov}}, \bibinfo {author} {\bibfnamefont {E.}~\bibnamefont {Schulz}}, \
  and\ \bibinfo {author} {\bibfnamefont {S.~J.}\ \bibnamefont {Gershman}},\
  }\href {\doibase 10.1038/s41562-020-01035-y} {\bibfield  {journal} {\bibinfo
  {journal} {Nature Human Behaviour}\ }\textbf {\bibinfo {volume} {5}},\
  \bibinfo {pages} {764} (\bibinfo {year} {2021})}\BibitemShut {NoStop}%
\bibitem [{\citenamefont {Zhang}\ \emph {et~al.}(2020)\citenamefont {Zhang},
  \citenamefont {Zhang}, \citenamefont {Chen},\ and\ \citenamefont
  {Liu}}]{ZhangOscillatory2020}%
  \BibitemOpen
  \bibfield  {author} {\bibinfo {author} {\bibfnamefont {S.}~\bibnamefont
  {Zhang}}, \bibinfo {author} {\bibfnamefont {J.}~\bibnamefont {Zhang}},
  \bibinfo {author} {\bibfnamefont {L.}~\bibnamefont {Chen}}, \ and\ \bibinfo
  {author} {\bibfnamefont {X.}~\bibnamefont {Liu}},\ }\href {\doibase
  10.1007/s11071-019-05398-4} {\bibfield  {journal} {\bibinfo  {journal}
  {Nonlinear Dynamics}\ }\textbf {\bibinfo {volume} {99}},\ \bibinfo {pages}
  {3301} (\bibinfo {year} {2020})}\BibitemShut {NoStop}%
\bibitem [{\citenamefont {Wang}\ \emph {et~al.}(2022)\citenamefont {Wang},
  \citenamefont {Jia}, \citenamefont {Zhang}, \citenamefont {Zhu},
  \citenamefont {Perc}, \citenamefont {Shi},\ and\ \citenamefont
  {Wang}}]{Wang2022Levy}%
  \BibitemOpen
  \bibfield  {author} {\bibinfo {author} {\bibfnamefont {L.}~\bibnamefont
  {Wang}}, \bibinfo {author} {\bibfnamefont {D.}~\bibnamefont {Jia}}, \bibinfo
  {author} {\bibfnamefont {L.}~\bibnamefont {Zhang}}, \bibinfo {author}
  {\bibfnamefont {P.}~\bibnamefont {Zhu}}, \bibinfo {author} {\bibfnamefont
  {M.}~\bibnamefont {Perc}}, \bibinfo {author} {\bibfnamefont {L.}~\bibnamefont
  {Shi}}, \ and\ \bibinfo {author} {\bibfnamefont {Z.}~\bibnamefont {Wang}},\
  }\href {\doibase 10.1007/s11071-022-07289-7} {\bibfield  {journal} {\bibinfo
  {journal} {Nonlinear Dynamics}\ }\textbf {\bibinfo {volume} {108}},\ \bibinfo
  {pages} {1837} (\bibinfo {year} {2022})}\BibitemShut {NoStop}%
\bibitem [{\citenamefont {Ding}\ \emph {et~al.}(2023)\citenamefont {Ding},
  \citenamefont {Zheng}, \citenamefont {Cai}, \citenamefont {Cai},
  \citenamefont {Chen}, \citenamefont {Zhang},\ and\ \citenamefont
  {Wang}}]{Ding2023Emergence}%
  \BibitemOpen
  \bibfield  {author} {\bibinfo {author} {\bibfnamefont {Z.}~\bibnamefont
  {Ding}}, \bibinfo {author} {\bibfnamefont {G.}~\bibnamefont {Zheng}},
  \bibinfo {author} {\bibfnamefont {C.}~\bibnamefont {Cai}}, \bibinfo {author}
  {\bibfnamefont {W.}~\bibnamefont {Cai}}, \bibinfo {author} {\bibfnamefont
  {L.}~\bibnamefont {Chen}}, \bibinfo {author} {\bibfnamefont {J.}~\bibnamefont
  {Zhang}}, \ and\ \bibinfo {author} {\bibfnamefont {X.}~\bibnamefont {Wang}},\
  }\href {\doibase https://doi.org/10.1016/j.chaos.2023.114032} {\bibfield
  {journal} {\bibinfo  {journal} {Chaos, Solitons \& Fractals}\ }\textbf
  {\bibinfo {volume} {175}},\ \bibinfo {pages} {114032} (\bibinfo {year}
  {2023})}\BibitemShut {NoStop}%
\bibitem [{\citenamefont {Zheng}\ \emph {et~al.}(2024)\citenamefont {Zheng},
  \citenamefont {Zhang}, \citenamefont {Deng}, \citenamefont {Cai},\ and\
  \citenamefont {Chen}}]{zheng2024evolution}%
  \BibitemOpen
  \bibfield  {author} {\bibinfo {author} {\bibfnamefont {G.}~\bibnamefont
  {Zheng}}, \bibinfo {author} {\bibfnamefont {J.}~\bibnamefont {Zhang}},
  \bibinfo {author} {\bibfnamefont {S.}~\bibnamefont {Deng}}, \bibinfo {author}
  {\bibfnamefont {W.}~\bibnamefont {Cai}}, \ and\ \bibinfo {author}
  {\bibfnamefont {L.}~\bibnamefont {Chen}},\ }\href {\doibase
  https://doi.org/10.1016/j.chaos.2024.115568} {\bibfield  {journal} {\bibinfo
  {journal} {Chaos, Solitons \& Fractals}\ }\textbf {\bibinfo {volume} {188}},\
  \bibinfo {pages} {115568} (\bibinfo {year} {2024})}\BibitemShut {NoStop}%
\bibitem [{\citenamefont {Jia}\ \emph {et~al.}(2021)\citenamefont {Jia},
  \citenamefont {Guo}, \citenamefont {Song}, \citenamefont {Shi}, \citenamefont
  {Deng}, \citenamefont {Perc},\ and\ \citenamefont {Wang}}]{JiaLocal2021}%
  \BibitemOpen
  \bibfield  {author} {\bibinfo {author} {\bibfnamefont {D.}~\bibnamefont
  {Jia}}, \bibinfo {author} {\bibfnamefont {H.}~\bibnamefont {Guo}}, \bibinfo
  {author} {\bibfnamefont {Z.}~\bibnamefont {Song}}, \bibinfo {author}
  {\bibfnamefont {L.}~\bibnamefont {Shi}}, \bibinfo {author} {\bibfnamefont
  {X.}~\bibnamefont {Deng}}, \bibinfo {author} {\bibfnamefont {M.}~\bibnamefont
  {Perc}}, \ and\ \bibinfo {author} {\bibfnamefont {Z.}~\bibnamefont {Wang}},\
  }\href {\doibase 10.1088/1367-2630/ac170a} {\bibfield  {journal} {\bibinfo
  {journal} {New Journal of Physics}\ }\textbf {\bibinfo {volume} {23}},\
  \bibinfo {pages} {083020} (\bibinfo {year} {2021})}\BibitemShut {NoStop}%
\bibitem [{\citenamefont {Zhao}\ \emph {et~al.}(2025)\citenamefont {Zhao},
  \citenamefont {Feng}, \citenamefont {Zheng}, \citenamefont {Cai},
  \citenamefont {Zhang},\ and\ \citenamefont {Chen}}]{ZhaoEvolution2025}%
  \BibitemOpen
  \bibfield  {author} {\bibinfo {author} {\bibfnamefont {C.}~\bibnamefont
  {Zhao}}, \bibinfo {author} {\bibfnamefont {X.}~\bibnamefont {Feng}}, \bibinfo
  {author} {\bibfnamefont {G.}~\bibnamefont {Zheng}}, \bibinfo {author}
  {\bibfnamefont {W.}~\bibnamefont {Cai}}, \bibinfo {author} {\bibfnamefont
  {J.}~\bibnamefont {Zhang}}, \ and\ \bibinfo {author} {\bibfnamefont
  {L.}~\bibnamefont {Chen}},\ }\href {\doibase 10.1103/4n16-56lf} {\bibfield
  {journal} {\bibinfo  {journal} {Phys. Rev. E}\ }\textbf {\bibinfo {volume}
  {112}},\ \bibinfo {pages} {054309} (\bibinfo {year} {2025})}\BibitemShut
  {NoStop}%
\bibitem [{\citenamefont {Zheng}\ \emph
  {et~al.}(2025{\natexlab{a}})\citenamefont {Zheng}, \citenamefont {Zhang},
  \citenamefont {Ou}, \citenamefont {Deng},\ and\ \citenamefont
  {Chen}}]{Zheng2024decoding}%
  \BibitemOpen
  \bibfield  {author} {\bibinfo {author} {\bibfnamefont {G.}~\bibnamefont
  {Zheng}}, \bibinfo {author} {\bibfnamefont {J.}~\bibnamefont {Zhang}},
  \bibinfo {author} {\bibfnamefont {X.}~\bibnamefont {Ou}}, \bibinfo {author}
  {\bibfnamefont {S.}~\bibnamefont {Deng}}, \ and\ \bibinfo {author}
  {\bibfnamefont {L.}~\bibnamefont {Chen}},\ }\href {\doibase
  10.1103/vk6m-48zs} {\bibfield  {journal} {\bibinfo  {journal} {Physical
  Review E}\ }\textbf {\bibinfo {volume} {111}},\ \bibinfo {pages} {064307}
  (\bibinfo {year} {2025}{\natexlab{a}})}\BibitemShut {NoStop}%
\bibitem [{\citenamefont {Zheng}\ \emph
  {et~al.}(2025{\natexlab{b}})\citenamefont {Zheng}, \citenamefont {Zhang},
  \citenamefont {Ou}, \citenamefont {Deng},\ and\ \citenamefont
  {Chen}}]{zheng2025decoding}%
  \BibitemOpen
  \bibfield  {author} {\bibinfo {author} {\bibfnamefont {G.}~\bibnamefont
  {Zheng}}, \bibinfo {author} {\bibfnamefont {J.}~\bibnamefont {Zhang}},
  \bibinfo {author} {\bibfnamefont {X.}~\bibnamefont {Ou}}, \bibinfo {author}
  {\bibfnamefont {S.}~\bibnamefont {Deng}}, \ and\ \bibinfo {author}
  {\bibfnamefont {L.}~\bibnamefont {Chen}},\ }\href {\doibase
  https://doi.org/10.1103/vk6m-48zs} {\bibfield  {journal} {\bibinfo  {journal}
  {Physical Review E}\ }\textbf {\bibinfo {volume} {111}},\ \bibinfo {pages}
  {064307} (\bibinfo {year} {2025}{\natexlab{b}})}\BibitemShut {NoStop}%
\bibitem [{\citenamefont {Zheng}\ \emph
  {et~al.}(2025{\natexlab{c}})\citenamefont {Zheng}, \citenamefont {Cai},
  \citenamefont {Qi}, \citenamefont {Zhang},\ and\ \citenamefont
  {Chen}}]{zheng2025optimal}%
  \BibitemOpen
  \bibfield  {author} {\bibinfo {author} {\bibfnamefont {G.}~\bibnamefont
  {Zheng}}, \bibinfo {author} {\bibfnamefont {W.}~\bibnamefont {Cai}}, \bibinfo
  {author} {\bibfnamefont {G.}~\bibnamefont {Qi}}, \bibinfo {author}
  {\bibfnamefont {J.}~\bibnamefont {Zhang}}, \ and\ \bibinfo {author}
  {\bibfnamefont {L.}~\bibnamefont {Chen}},\ }\href {\doibase
  https://doi.org/10.1103/z71v-xmxk} {\bibfield  {journal} {\bibinfo  {journal}
  {Physical Review E}\ }\textbf {\bibinfo {volume} {112}},\ \bibinfo {pages}
  {064305} (\bibinfo {year} {2025}{\natexlab{c}})}\BibitemShut {NoStop}%
\bibitem [{\citenamefont {Quan}\ \emph {et~al.}(2020)\citenamefont {Quan},
  \citenamefont {Tang}, \citenamefont {Zhou}, \citenamefont {Wang},\ and\
  \citenamefont {Yang}}]{Quan2020Reputation}%
  \BibitemOpen
  \bibfield  {author} {\bibinfo {author} {\bibfnamefont {J.}~\bibnamefont
  {Quan}}, \bibinfo {author} {\bibfnamefont {C.}~\bibnamefont {Tang}}, \bibinfo
  {author} {\bibfnamefont {Y.}~\bibnamefont {Zhou}}, \bibinfo {author}
  {\bibfnamefont {X.}~\bibnamefont {Wang}}, \ and\ \bibinfo {author}
  {\bibfnamefont {J.-B.}\ \bibnamefont {Yang}},\ }\href {\doibase
  10.1016/j.chaos.2019.109517} {\bibfield  {journal} {\bibinfo  {journal}
  {Chaos, Solitons \& Fractals}\ }\textbf {\bibinfo {volume} {131}},\ \bibinfo
  {pages} {109517} (\bibinfo {year} {2020})}\BibitemShut {NoStop}%
\bibitem [{\citenamefont {Quan}\ \emph {et~al.}(2021)\citenamefont {Quan},
  \citenamefont {Tang},\ and\ \citenamefont {Wang}}]{QUAN2021Reputation}%
  \BibitemOpen
  \bibfield  {author} {\bibinfo {author} {\bibfnamefont {J.}~\bibnamefont
  {Quan}}, \bibinfo {author} {\bibfnamefont {C.}~\bibnamefont {Tang}}, \ and\
  \bibinfo {author} {\bibfnamefont {X.}~\bibnamefont {Wang}},\ }\href {\doibase
  https://doi.org/10.1016/j.physa.2020.125488} {\bibfield  {journal} {\bibinfo
  {journal} {Physica A: Statistical Mechanics and its Applications}\ }\textbf
  {\bibinfo {volume} {563}},\ \bibinfo {pages} {125488} (\bibinfo {year}
  {2021})}\BibitemShut {NoStop}%
\bibitem [{\citenamefont {Quan}\ \emph {et~al.}(2024)\citenamefont {Quan},
  \citenamefont {Zhang}, \citenamefont {Chen}, \citenamefont {Tang},\ and\
  \citenamefont {Wang}}]{QUAN2024Reputation}%
  \BibitemOpen
  \bibfield  {author} {\bibinfo {author} {\bibfnamefont {J.}~\bibnamefont
  {Quan}}, \bibinfo {author} {\bibfnamefont {X.}~\bibnamefont {Zhang}},
  \bibinfo {author} {\bibfnamefont {W.}~\bibnamefont {Chen}}, \bibinfo {author}
  {\bibfnamefont {C.}~\bibnamefont {Tang}}, \ and\ \bibinfo {author}
  {\bibfnamefont {X.}~\bibnamefont {Wang}},\ }\href {\doibase
  https://doi.org/10.1016/j.amc.2024.128745} {\bibfield  {journal} {\bibinfo
  {journal} {Applied Mathematics and Computation}\ }\textbf {\bibinfo {volume}
  {475}},\ \bibinfo {pages} {128745} (\bibinfo {year} {2024})}\BibitemShut
  {NoStop}%
\bibitem [{\citenamefont {Feng}\ \emph {et~al.}(2024)\citenamefont {Feng},
  \citenamefont {Han}, \citenamefont {Feng},\ and\ \citenamefont
  {Szolnoki}}]{Feng2024An}%
  \BibitemOpen
  \bibfield  {author} {\bibinfo {author} {\bibfnamefont {K.}~\bibnamefont
  {Feng}}, \bibinfo {author} {\bibfnamefont {S.}~\bibnamefont {Han}}, \bibinfo
  {author} {\bibfnamefont {M.}~\bibnamefont {Feng}}, \ and\ \bibinfo {author}
  {\bibfnamefont {A.}~\bibnamefont {Szolnoki}},\ }\href {\doibase
  10.1016/j.amc.2024.128618} {\bibfield  {journal} {\bibinfo  {journal}
  {Applied Mathematics and Computation}\ }\textbf {\bibinfo {volume} {472}},\
  \bibinfo {pages} {128618} (\bibinfo {year} {2024})}\BibitemShut {NoStop}%
\bibitem [{\citenamefont {Zhang}\ \emph {et~al.}(2024)\citenamefont {Zhang},
  \citenamefont {Liu},\ and\ \citenamefont {Zhang}}]{zhang2024Reputation}%
  \BibitemOpen
  \bibfield  {author} {\bibinfo {author} {\bibfnamefont {Q.}~\bibnamefont
  {Zhang}}, \bibinfo {author} {\bibfnamefont {J.}~\bibnamefont {Liu}}, \ and\
  \bibinfo {author} {\bibfnamefont {X.}~\bibnamefont {Zhang}},\ }\href
  {\doibase https://doi.org/10.1016/j.physa.2024.129999} {\bibfield  {journal}
  {\bibinfo  {journal} {Physica A: Statistical Mechanics and its Applications}\
  }\textbf {\bibinfo {volume} {650}},\ \bibinfo {pages} {129999} (\bibinfo
  {year} {2024})}\BibitemShut {NoStop}%
\bibitem [{\citenamefont {Zou}\ and\ \citenamefont
  {Huang}(2024)}]{Zou2024Incorporating}%
  \BibitemOpen
  \bibfield  {author} {\bibinfo {author} {\bibfnamefont {K.}~\bibnamefont
  {Zou}}\ and\ \bibinfo {author} {\bibfnamefont {C.}~\bibnamefont {Huang}},\
  }\href {\doibase https://doi.org/10.1016/j.chaos.2024.115203} {\bibfield
  {journal} {\bibinfo  {journal} {Chaos, Solitons \& Fractals}\ }\textbf
  {\bibinfo {volume} {186}},\ \bibinfo {pages} {115203} (\bibinfo {year}
  {2024})}\BibitemShut {NoStop}%
\bibitem [{\citenamefont {Ren}\ and\ \citenamefont
  {Zeng}(2023)}]{ren2023reputation}%
  \BibitemOpen
  \bibfield  {author} {\bibinfo {author} {\bibfnamefont {T.}~\bibnamefont
  {Ren}}\ and\ \bibinfo {author} {\bibfnamefont {X.-J.}\ \bibnamefont {Zeng}},\
  }\href {\doibase 10.1109/TEVC.2023.3304911} {\bibfield  {journal} {\bibinfo
  {journal} {IEEE Transactions on Evolutionary Computation}\ }\textbf {\bibinfo
  {volume} {28}},\ \bibinfo {pages} {1177} (\bibinfo {year}
  {2023})}\BibitemShut {NoStop}%
\bibitem [{\citenamefont {Hu}\ \emph {et~al.}(2024)\citenamefont {Hu},
  \citenamefont {Jin}, \citenamefont {Jiang},\ and\ \citenamefont
  {Liu}}]{Hu2024Reputation}%
  \BibitemOpen
  \bibfield  {author} {\bibinfo {author} {\bibfnamefont {Q.}~\bibnamefont
  {Hu}}, \bibinfo {author} {\bibfnamefont {T.}~\bibnamefont {Jin}}, \bibinfo
  {author} {\bibfnamefont {Y.}~\bibnamefont {Jiang}}, \ and\ \bibinfo {author}
  {\bibfnamefont {X.}~\bibnamefont {Liu}},\ }\href {\doibase
  https://doi.org/10.1016/j.amc.2023.128445} {\bibfield  {journal} {\bibinfo
  {journal} {Applied Mathematics and Computation}\ }\textbf {\bibinfo {volume}
  {466}},\ \bibinfo {pages} {128445} (\bibinfo {year} {2024})}\BibitemShut
  {NoStop}%
\bibitem [{\citenamefont {Fang}\ \emph {et~al.}(2025)\citenamefont {Fang},
  \citenamefont {Xu}, \citenamefont {Xie}, \citenamefont {Yue}, \citenamefont
  {Benko},\ and\ \citenamefont {Huang}}]{Fang2025Evolution}%
  \BibitemOpen
  \bibfield  {author} {\bibinfo {author} {\bibfnamefont {Z.}~\bibnamefont
  {Fang}}, \bibinfo {author} {\bibfnamefont {H.}~\bibnamefont {Xu}}, \bibinfo
  {author} {\bibfnamefont {C.}~\bibnamefont {Xie}}, \bibinfo {author}
  {\bibfnamefont {X.}~\bibnamefont {Yue}}, \bibinfo {author} {\bibfnamefont
  {T.~P.}\ \bibnamefont {Benko}}, \ and\ \bibinfo {author} {\bibfnamefont
  {C.}~\bibnamefont {Huang}},\ }\href {\doibase
  https://doi.org/10.1016/j.chaos.2025.117115} {\bibfield  {journal} {\bibinfo
  {journal} {Chaos, Solitons \& Fractals}\ }\textbf {\bibinfo {volume} {200}},\
  \bibinfo {pages} {117115} (\bibinfo {year} {2025})}\BibitemShut {NoStop}%
\bibitem [{\citenamefont {Cuesta}\ \emph {et~al.}(2015)\citenamefont {Cuesta},
  \citenamefont {Gracia-L{\'a}zaro}, \citenamefont {Ferrer}, \citenamefont
  {Moreno},\ and\ \citenamefont {S{\'a}nchez}}]{cuesta2015reputation}%
  \BibitemOpen
  \bibfield  {author} {\bibinfo {author} {\bibfnamefont {J.~A.}\ \bibnamefont
  {Cuesta}}, \bibinfo {author} {\bibfnamefont {C.}~\bibnamefont
  {Gracia-L{\'a}zaro}}, \bibinfo {author} {\bibfnamefont {A.}~\bibnamefont
  {Ferrer}}, \bibinfo {author} {\bibfnamefont {Y.}~\bibnamefont {Moreno}}, \
  and\ \bibinfo {author} {\bibfnamefont {A.}~\bibnamefont {S{\'a}nchez}},\
  }\href {\doibase https://doi.org/10.1038/srep07843} {\bibfield  {journal}
  {\bibinfo  {journal} {Scientific reports}\ }\textbf {\bibinfo {volume} {5}},\
  \bibinfo {pages} {7843} (\bibinfo {year} {2015})}\BibitemShut {NoStop}%
\bibitem [{\citenamefont {Molleman}\ \emph {et~al.}(2014)\citenamefont
  {Molleman}, \citenamefont {Van~den Berg},\ and\ \citenamefont
  {Weissing}}]{molleman2014consistent}%
  \BibitemOpen
  \bibfield  {author} {\bibinfo {author} {\bibfnamefont {L.}~\bibnamefont
  {Molleman}}, \bibinfo {author} {\bibfnamefont {P.}~\bibnamefont {Van~den
  Berg}}, \ and\ \bibinfo {author} {\bibfnamefont {F.~J.}\ \bibnamefont
  {Weissing}},\ }\href {\doibase https://doi.org/10.1038/ncomms4570} {\bibfield
   {journal} {\bibinfo  {journal} {Nature communications}\ }\textbf {\bibinfo
  {volume} {5}},\ \bibinfo {pages} {3570} (\bibinfo {year} {2014})}\BibitemShut
  {NoStop}%
\bibitem [{\citenamefont {Yang}\ \emph {et~al.}(2024)\citenamefont {Yang},
  \citenamefont {Zheng}, \citenamefont {Perc},\ and\ \citenamefont
  {Li}}]{YANG2024Interaction}%
  \BibitemOpen
  \bibfield  {author} {\bibinfo {author} {\bibfnamefont {Z.}~\bibnamefont
  {Yang}}, \bibinfo {author} {\bibfnamefont {L.}~\bibnamefont {Zheng}},
  \bibinfo {author} {\bibfnamefont {M.}~\bibnamefont {Perc}}, \ and\ \bibinfo
  {author} {\bibfnamefont {Y.}~\bibnamefont {Li}},\ }\href {\doibase
  https://doi.org/10.1016/j.amc.2023.128364} {\bibfield  {journal} {\bibinfo
  {journal} {Applied Mathematics and Computation}\ }\textbf {\bibinfo {volume}
  {463}},\ \bibinfo {pages} {128364} (\bibinfo {year} {2024})}\BibitemShut
  {NoStop}%
\bibitem [{\citenamefont {Lamba}(2014)}]{Lamba2014Social}%
  \BibitemOpen
  \bibfield  {author} {\bibinfo {author} {\bibfnamefont {S.}~\bibnamefont
  {Lamba}},\ }\href {\doibase 10.1098/rspb.2014.0417} {\bibfield  {journal}
  {\bibinfo  {journal} {Proceedings of the Royal Society B: Biological
  Sciences}\ }\textbf {\bibinfo {volume} {281}},\ \bibinfo {pages} {20140417}
  (\bibinfo {year} {2014})}\BibitemShut {NoStop}%
\bibitem [{\citenamefont {Sheng}\ \emph {et~al.}(2024)\citenamefont {Sheng},
  \citenamefont {Zhang}, \citenamefont {Zheng}, \citenamefont {Zhang},
  \citenamefont {Cai},\ and\ \citenamefont {Chen}}]{shengCatalytic2024}%
  \BibitemOpen
  \bibfield  {author} {\bibinfo {author} {\bibfnamefont {A.}~\bibnamefont
  {Sheng}}, \bibinfo {author} {\bibfnamefont {J.}~\bibnamefont {Zhang}},
  \bibinfo {author} {\bibfnamefont {G.}~\bibnamefont {Zheng}}, \bibinfo
  {author} {\bibfnamefont {J.}~\bibnamefont {Zhang}}, \bibinfo {author}
  {\bibfnamefont {W.}~\bibnamefont {Cai}}, \ and\ \bibinfo {author}
  {\bibfnamefont {L.}~\bibnamefont {Chen}},\ }\href {\doibase
  10.1063/5.0231772} {\bibfield  {journal} {\bibinfo  {journal} {Chaos: An
  Interdisciplinary Journal of Nonlinear Science}\ }\textbf {\bibinfo {volume}
  {34}},\ \bibinfo {pages} {103117} (\bibinfo {year} {2024})}\BibitemShut
  {NoStop}%
\end{thebibliography}%
	\end{document}